\documentclass[12pt,draft]{article}
\usepackage[british]{babel}
\def\AnswerYes{y}
\def\draftVersion{n}               
\def\feynVersion{n}                
\usepackage[T1]{fontenc}

\usepackage{cite}                         
\usepackage{citesort}                     
\usepackage[dvips]{color}                 
\usepackage[final]{graphicx}              
\graphicspath{{figures/}{./}}

\usepackage{multirow}                     
\usepackage{amssymb}
\usepackage{amsmath}
\usepackage{xspace}                       
\ifx\feynVersion\AnswerYes
   \usepackage{feynmp}                    
\fi

\ifx\draftVersion\AnswerYes
   \usepackage[scriptsize]{myshowkeys}    
   \newcommand{\comment}[1]{%
     {\scriptsize \sffamily \bfseries #1} }
   \newcommand{\margin}[1]{%
     \marginpar{\scriptsize \sffamily \bfseries #1} }
   \newcommand{\drafty}{\textbf{Draft version \today} \hfill}
\else
   \newcommand{\comment}[1]{}
   \newcommand{\margin}[1]{}
   \newcommand{\drafty}{}
\fi
\newcommand{\absatz}{\vspace{2ex}\noindent}

\newcommand{\journal}[4]{\textit{#1}\xspace\textbf{#2}, #4 (#3)}

\newcommand{\NPA}{\textnormal{Nucl.\ Phys.\ }\textnormal{A}}
\newcommand{\NPB}{\textnormal{Nucl.\ Phys.\ }\textnormal{B}}
\newcommand{\PLB}{\textnormal{Phys.\ Lett.\ }\textnormal{B}}
\newcommand{\PR}{\textnormal{Phys.\ Rev.\ }}

\newcommand{\PRC}{\PR\textnormal{C}~}
\newcommand{\PRD}{\PR\textnormal{D}~}
\newcommand{\PRL}{\PR\textnormal{Lett.\ }}

\newcommand{\dis}{\displaystyle}

\newcommand{\non}{\nonumber}

\newcommand{\half}{\frac{1}{2}}

\newcommand{\ii}{\mathrm{i}}
\newcommand{\dd}{\mathrm{d}}

\newcommand{\deint}[2]{\dd^{#1}\! #2\;}

\newcommand{\kv}{\vec{k}}
\newcommand{\pv}{\vec{p}}
\newcommand{\qv}{\vec{q}}

\newcommand{\mpi}{\ensuremath{m_\pi}}

\newcommand{\MeV}{\ensuremath{\mathrm{MeV}}}
\newcommand{\fm}{\ensuremath{\mathrm{fm}}}
\newcommand{\EFTNoPion}{EFT(${\pi\hskip-0.55em /}$)\xspace}

\newcommand{\NXLO}[1]{N\ensuremath{{}^{#1}}LO\xspace}

\newcommand{\wave}[3]{\ensuremath{{}^{#1}\mathrm{#2}_{#3}}}
\newcommand{\oneS}{\wave{1}{S}{0}}
\newcommand{\twoS}{\wave{2}{S}{\half}}
\newcommand{\threeS}{\wave{3}{S}{1}}
\newcommand{\fourS}{\wave{4}{S}{\frac{3}{2}}}

\newcommand{\LambdaNoPion}{\ensuremath{\Lambda_{\pi\hskip-0.4em /}}}

\renewcommand{\Re}{\mathrm{Re}}
\renewcommand{\Im}{\mathrm{Im}}

 \newcommand{\calK}{\mathcal{K}}
 \newcommand{\calM}{\mathcal{M}}

\newcommand{\mytitle}[1]{\begin{center}\LARGE{\textbf{#1}}\end{center}}
\newcommand{\myauthor}[1]{\textbf{#1}}
\newcommand{\myaddress}[1]{\textit{#1}}
\newcommand{\mypreprint}[1]{\begin{flushright}#1\end{flushright}}

\begin{document}
%

\begin{titlepage}
  \setcounter{page}{0} \mypreprint{
    \drafty
    nucl-th/0502039\\
    TUM-T39-04-20\\
    13th February 2005 \\
    Revised version 4th March 2005\\
    Re-Revised version 27th May 2005\\
    Final version 1st July 2005\\
    Accepted by Nuclear Physics \textbf{A}
  }
  
  
  \mytitle{Na\"ive Dimensional Analysis for \\
    Three-Body Forces Without Pions}
  

\begin{center}
  \myauthor{Harald W.\ Grie\3hammer$^{a,}$}\footnote{Email:
    hgrie@physik.tu-muenchen.de;
    permanent address: a}\\[2ex]
  
  \vspace*{0.5cm}
  
  \myaddress{$^a$
    Institut f{\"u}r Theoretische Physik (T39), Physik-Department,\\
    Technische Universit{\"a}t M{\"u}nchen, D-85747 Garching, Germany}
  
  \vspace*{0.2cm}

\end{center}


\begin{abstract}
  For systems of three identical particles in which short-range forces produce
  shallow two-particle bound states, and in particular for the ``pion-less''
  Effective Field Theory of Nuclear Physics, I extend and systematise the
  power-counting of three-body forces to all partial waves and orders,
  including external currents. With low-energy observables independent of the
  details of short-distance dynamics, the typical strength of a three-body
  force is determined from the superficial degree of divergence of the
  three-body diagrams which contain only two-body forces. This na\"ive
  dimensional analysis must be amended as the asymptotic solution to the
  leading-order Faddeev equation depends for large off-shell momenta crucially
  on the partial wave and spin-combination of the system. It is shown by
  analytic construction to be weaker than expected in most channels with
  angular momentum smaller than $3$. This demotes many three-nucleon forces to
  high orders. Observables like the \fourS-scattering length are less
  sensitive to three-nucleon forces than guessed.
  I also comment on the Efimov effect and limit-cycle for non-zero angular
  momentum.
\end{abstract}
\vskip 1.0cm
\noindent
\begin{tabular}{rl}
Suggested PACS numbers:& \begin{minipage}[t][\height][t]{10.7cm}
                    02.30.Rz, 02.30.Uu, 11.80.Jy, 13.75.Cs, 14.20.Dh,
                    21.30.-x, 25.40.Dn, 27.10.+h  
                    \end{minipage}
                    \\[4ex]
Suggested Keywords: &\begin{minipage}[t]{10.7cm}
                    Effective Field Theory, three-body system,
                    three-body force, Faddeev equation, partial waves,
                    Mellin transform. 
                    \end{minipage}
\end{tabular}

\vskip 1.0cm

\end{titlepage}

\setcounter{footnote}{0}

\newpage

%

\section{Introduction}
\setcounter{equation}{0}
\label{sec:introduction}

Three-body forces parameterise the interactions between three particles on
scales much smaller than what can be resolved by two-body interactions.
Traditionally, they were often introduced \emph{a posteriori} to cure
discrepancies between experiment and theory, but such an approach is of course
untenable when data are scarce or absent, predictive power is required, or
one- or two-body properties are extracted from three-body data.

The pivotal promise of an Effective Field Theory (EFT) is that it describes
all Physics below a certain ``breakdown-scale'' to a given accuracy with the
minimal set of parameters, and that it hence predicts in a model-independent
way the typical strength with which also three-body forces enter in
observables. This promise is based on a dimension-less, small parameter: the
typical momentum of the low-energy process in units of the breakdown-scale,
namely of the scale on which details of the short-range interactions are
resolved. It allows one to truncate the momentum-expansion of the forces at a
given level of accuracy, keeping only and all the terms up to a given order,
and thus establishes a ``power-counting'' of all forces. One can then estimate
\emph{a priori} the experimental accuracy necessary to dis-entangle particular
effects like a three-body force in observables.

The power-counting is to a high degree determined by na\"ive dimensional
analysis~\cite{NDA}. As low-energy observables must be insensitive to the
details of short-distance Physics, they are in particular independent of a
cut-off $\Lambda$ employed to regulate the theory at short distances.
Typically, a divergence from loop integrations must therefore be cancelled by
at least one coefficient $C(\Lambda)$ in the EFT Lagrangean, which thus also
encodes short-distance dynamics. This counter-term enters hence at the same
order as the first divergence which it must absorb. It ensures that the EFT is
cut-off independent at each order, and therefore renormalisable and
self-consistent. With the running of the counter-term thus determined, its
initial condition provides an unknown, free parameter which has to be found
from experiment. Na\"ive dimensional analysis assumes now that the typical
size of the counter-term is ``natural'', i.e.~at most of the same magnitude as
the size of its running: $C(\Lambda)\sim C(2\Lambda)\sim
C(2\Lambda)-C(\Lambda)$.  Thus, it guarantees that the EFT contains at a given
order the minimal number of free parameters which are necessary to render the
theory renormalisable, and by that also the minimal number of independent
low-energy coefficients necessary to describe all low-energy phenomenology to
a given level of accuracy.

When all interactions are treated perturbatively, like in Chiral Perturbation
Theory in the purely mesonic sector, na\"ive dimensional analysis amounts to
little more than counting the mass-dimension of an interaction~\cite{NDA}. I
will demonstrate that such reasoning becomes however too simplistic when some
interactions in the EFT must be treated non-perturbatively. This is the case
in Nuclear Physics, and for some systems of Atomic Physics.

\absatz While the separation of scales between low-energy and high-energy
degrees of freedom in Nuclear Physics makes it an ideal playground for
EFT-methods, finding such a power-counting proves a difficult goal for
few-nucleon systems. To establish a formalism which is self-consistent, agrees
with nuclear phenomenology and can firmly be rooted in QCD, one has to cope
with shallow real and virtual few-nucleon bound-states. The deuteron size of
$\approx 4.3\;\fm$ for example seems not to be connected even to the soft
scales of QCD, e.g.~the pion mass or decay constant. The effective low-energy
degrees of freedom and symmetries of QCD dictate a unique Lagrangean; but
there are conceptually quite different ordering-schemes available for the
few-nucleon system which lead to different, experimentally falsifiable
predictions. The na\"ivest versions perturb around the free theory and hence
cannot accommodate shallow bound-states at all. They are self-consistent, but
not consistent with Nature. Weinberg~\cite{Weinberg} proposed to build
few-nucleon systems from a nucleon-nucleon potential which consists of contact
interactions and pion-exchanges constrained by chiral symmetry. The
interactions in the potential are ordered following na\"ive dimensional
analysis as if the theory would be perturbative, and the potential is then
iterated to produce the unnaturally large length scales in the two-nucleon
system by fine-tuning between long- and short-distance contributions.
Few-nucleon interactions are added using the same prescription. Whether this
approach correctly and self-consistently reproduces QCD at low energies in the
three- and more-nucleon sector is an open question.

However, low-lying few-body bound-states also offer the opportunity for a more
radical approach: For momenta below the pion-mass, the only forces can be
taken to be point-like two- and more-nucleon interactions. This Nuclear
Effective Field Theory with pions integrated out (\EFTNoPion) is in the
two-nucleon system manifestly self-consistent and proves -- on quite general
grounds -- to be the correct version of QCD at extremely low energies, once
fine-tuning is observed, see
Refs.~\cite{bira_review,seattle_review,bedaque_bira_review} for recent
reviews. A plethora of pivotal physical processes which are both interesting
in their own right and important for astrophysical applications and
fundamental questions, e.g.~big-bang nucleo-synthesis and static neutron
properties, were investigated with high accuracy. One obtains usually quite
simple, analytic results, and most of the coefficients are determined by
simple, well-known long-range observables. Recently, a manifestly
self-consistent power-counting for the three-nucleon forces of \EFTNoPion in
the \twoS-wave of $Nd$-scattering was
established~\cite{3stooges_doublet,doubletNLO,4stooges}. First high-accuracy
calculations also including external sources are now
performed~\cite{Sadeghi:2004es}. A remarkable phenomenon of this channel is
that the first three-body force appears already at leading order to stabilise
the wave-function against collapse~\cite{danilov}, leading to a new
renormalisation-group phenomenon, the
``limit-cycle''~\cite{3stooges_boson,wilson,Braaten:2003eu}, manifested also
by the Efimov effect~\cite{efimovI,3stooges_boson}.  This can also be shown
using a subtraction method~\cite{Afnan:2003bs}. It was also confirmed by an
analysis of the renormalisation-group flow in the position-space version of
the problem~\cite{BarfordBirse}.

\EFTNoPion is universal in a dual sense: First, its methods can be applied to
a host of systems in which short-range forces conspire to produce shallow
two-particle bound states: One example are identical spin-less bosons, found
in bound-states of neutral atoms like the ${}^4$He-dimer and -trimer which are
bound by van-der-Waals-forces, or loss rates in Bose-Einstein condensates near
Feshbach resonances, see Ref.~\cite{Braaten:2004rn} for a review. Our results
are thus readily taken over to such systems. Second, any consistent EFT of
nucleons and pions must reduce to \EFTNoPion in the extreme low-energy limit.
Therefore, lessons learned in the latter shed light on the consistent
systematisation of the former. 
As EFTs are model-independent, considerably more sophisticated and
computationally involved potential-model calculations must agree with the
predictions of \EFTNoPion when they reproduce the two- and three-body data
which are used as input for \EFTNoPion to the same level of accuracy.

\absatz This article confronts the power-counting of three-body forces in any
three-particle system with large two-particle scattering lengthes and only
contact interactions. It is organised as follows: In
Section~\ref{sec:formalism}, the necessary foundations are summarised. After
establishing the superficial degree of divergence of diagrams which contain
only two-body forces in Sect~\ref{sec:higherorders}, the far-offshell
amplitude of the leading-order Faddeev equation in each partial wave is
determined analytically in Sect.~\ref{sec:ordering} together with a short
discussion of the Efimov effect in non-integer partial waves.
Section~\ref{sec:ordering} then classifies at which order a given three-body
force is needed to render cut-off independent results. Physically relevant
consequences are discussed in Sect.~\ref{sec:consequences}, together with some
caveats. 
After the Conclusions, an Appendix sketches some mathematical details. I also
correct some errors in a brief summary of some of the results in
Ref.~\cite{suppressed3bfs}.

\section{Three-Body Forces in \EFTNoPion} \setcounter{equation}{0}
\label{sec:3NFs}

\subsection{Three-Body Systems with Large Two-Body Scattering-Length} 
\label{sec:formalism}

\begin{figure}[!htb]
\begin{center}
  \includegraphics*[width=0.82\textwidth]{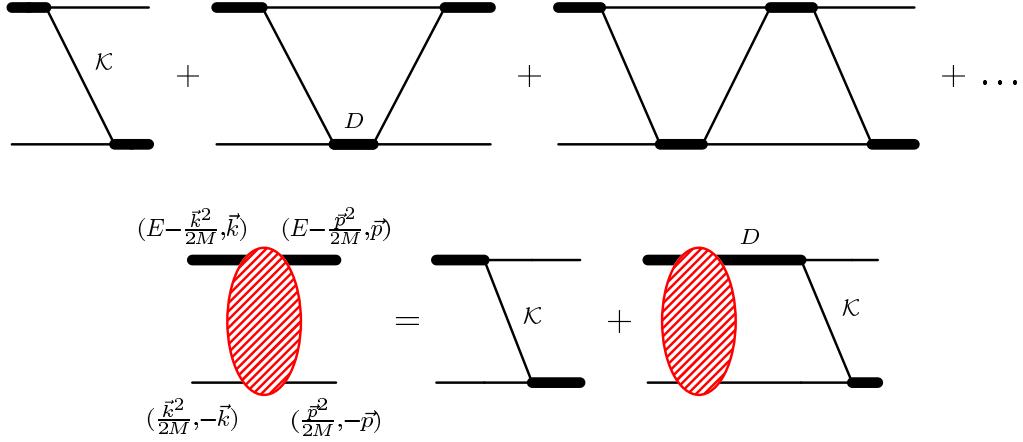}
\caption{Re-summation of the infinite number of LO three-body diagrams (top)
  into the corresponding Faddeev integral equation (bottom). Thick line ($D$):
  two-nucleon propagator; thin line ($\calK$): propagator of the exchanged
  nucleon; ellipse: LO half off-shell amplitude. }
\label{fig:kinematics}
\end{center}
\end{figure}

We consider three identical particles $N$ of mass $M$ interacting only with
contact forces such that two particles form a shallow real or virtual two-body
bound-state $d$. As the steps leading to the leading-order (LO)
scattering-amplitude $dN\to dN$ were often described in the literature, they
are not covered here; see Ref.~\cite{improve3body} also for the notation used
in the following. For convenience, a ``deuteron'' field is introduced as the
auxiliary field which describes scattering between two particles with an
anomalously large scattering length
$1/\gamma$~\cite{Kaplan:1996nv,2stooges_quartet,3stooges_quartet,pbhg}. Its
propagator is therefore given by the LO-truncation of the effective-range
expansion~\cite{Bethe}:
\begin{equation}
  \label{eq:dprop}
D(q_0,\qv)=\frac{1}{\gamma-\sqrt{\frac{\qv^2}{4}-Mq_0-\ii\epsilon}}
\end{equation}
A real bound-state $d$ has at this order the binding energy
$\gamma^2/M\ll\LambdaNoPion$, much smaller than the breakdown-scale of EFT on
which new degrees of freedom are resolved. In the three-body system, an
infinite number of diagrams contributes at LO, see Fig.~\ref{fig:kinematics}.
The corresponding Faddeev equation for scattering between the auxiliary field
$d$ and the remaining particle, first derived by Skorniakov and
Ter-Martirosian~\cite{skorny}, is unitarily equivalent to the original problem
of scattering between three particles~\cite{pbhg,chickenpaper}. One finds for
$Nd$-scattering in the $l$th partial wave in the centre-of-mass system the
integral equation for the half off-shell amplitude (before wave-function
renormalisation)
\begin{eqnarray}
  \label{eq:faddeev}
  t^{(l)}_\lambda(E;k,p)=8\pi\lambda\; \calK^{(l)}(E;k,p)
  -\frac{4}{\pi}\;\lambda
  \int\limits_0^\infty\deint{}{q} q^2\;\calK^{(l)}(E;q,p)\;
  D(E-\frac{q^2}{2M},q)\;t^{(l)}_\lambda(E;k,q)\;.
\end{eqnarray}
With the kinematics defined in Fig.~\ref{fig:kinematics},
$E:=\frac{3\kv^2}{4M}-\frac{\gamma^2}{M}-\ii\epsilon$ is the total
non-relativistic energy; $\kv$ the relative momentum of the incoming deuteron;
$\pv$ the off-shell momentum of the outgoing one, with $p=k$ the on-shell
point; and the projected propagator of the exchanged particle on angular
momentum $l$ is
\begin{equation}
  \label{eq:projectedNpropagator}
  \calK^{(l)}(E;q,p):=\frac{1}{2}\;\int\limits_{-1}^1\deint{}{x}
  \frac{P_l(x)}{p^2+q^2-ME+ pqx}
  =\frac{(-1)^l}{pq}\;Q_l\left(\frac{p^2+q^2-ME}{pq}\right).
\end{equation}
The $l$th Legendre polynomial of the second kind with complex argument is
defined as in~\cite{Gradstein}
\begin{equation}
  \label{eq:legendreQ}
   Q_l(z)=\half\int\limits_{-1}^1\deint{}{t}\frac{P_l(t)}{z-t}\;\;.
\end{equation}
The ``spin-parameter'' $\lambda$ depends on the spins of the three particles
and how they combine. The values for the physically most relevant three-body
systems are summarised in Table~\ref{tab:lambda}.
\begin{table}[!htb]
  \centering
  \begin{tabular}{|c|c|}
    \hline
   $\lambda=1$&$\lambda=-\half$\\
   \hline
   3 spin-less bosons & 3 nucleons coupled to $S=\frac{3}{2}$\\
   \hline
   Wigner-symmetric part of & Wigner-antisymmetric part of \\
   3 nucleons coupled to $S=\half$&
   3 nucleons coupled to $S=\half$\\
   \hline
  \end{tabular}
  \caption{The spin-parameter $\lambda$ for physical systems of identical
    particles described by \eqref{eq:faddeev}.} 
  \label{tab:lambda}
\end{table}

In \EFTNoPion, $Nd$-scattering in the $S=\half$-channel is at first sight
described by a more complex integral equation because two-nucleon scattering
has two anomalously large scattering lengthes: $1/\gamma_s$ in the
\oneS-channel, and $1/\gamma_t$ in the \threeS-channel. Therefore, two
cluster-configurations exist in the three-nucleon system: In one, the
spin-triplet auxiliary field $d_t$ (the deuteron) combines with the
``spectator'' nucleon $N$ to total spin $S=\frac{3}{2}$ (quartet channel) or
$S=\frac{1}{2}$ (doublet channel), depending on whether the deuteron and
nucleon spins are parallel or anti-parallel. In the other, the spin-singlet
auxiliary field $d_s$ combines with the remaining nucleon to total spin
$S=\frac{1}{2}$. In the doublet channel, the Faddeev equation is thus
two-dimensional: The amplitude $t^{(l)}_{d,tt}$ stands for the $Nd_t\to
Nd_t$-process, $t^{(l)}_{d,ts}$ for the $Nd_t\to Nd_s$-process, and with
$D_{s/t}$ defined analogously to \eqref{eq:dprop}:
\begin{eqnarray}
  \label{eq:doubletpw}
  \lefteqn{{t^{(l)}_{d,tt}\choose t^{(l)}_{d,ts}}(E;k,p)=2\pi\;
  \calK^{(l)}(E;k,p)\;{1\choose -3}}\\
  &&
  -\;\frac{1}{\pi}\int\limits_0^\infty\deint{}{q} q^2\;
    \calK^{(l)}(E;q,p)\;\begin{pmatrix}1&-3\\-3&1\end{pmatrix}
    \begin{pmatrix}D_t(E-\frac{q^2}{2M},p)&0\\0&D_s(E-\frac{q^2}{2M},p)
    \end{pmatrix}
    {t^{(l)}_{d,tt}\choose t^{(l)}_{d,ts}}(E;k,q)
  \non
\end{eqnarray}
In the following, we are only interested in the unphysical short-distance
behaviour of the amplitudes, i.e.~in the UV-limit for the half off-shell
momenta of (\ref{eq:doubletpw}): $p,q\gg k,\;E,\;\gamma_{s/t}$. This suffices
to determine in Sect.~\ref{sec:ordering} the order at which divergences need
to be cancelled by counter-terms parameterising three-nucleon interactions. In
this limit, the $NN$ scattering-amplitudes are automatically
Wigner-$SU(4)$-symmetric, i.e.~symmetric under arbitrary combined rotations of
spin and iso-spin~\cite{su4,Mehen:1999qs,3stooges_doublet}:
\begin{equation}
  \label{eq:dforWigner}\lim\limits_{q\gg E,\gamma_{s/t}}
  D_{s/t}(E-\frac{\qv^2}{2M},\qv)=\lim\limits_{q\gg E,\gamma}
  D(E-\frac{\qv^2}{2M},\qv)=-\frac{2}{\sqrt{3}} \;\frac{1}{q}
\end{equation}
Building the following linear combinations which are symmetric
resp.~anti-symmetric under Wigner-transformations,
\begin{equation}
  \label{eq:lincombforWigner}
  t_{\text{Ws}}^{(l)}:=\half\left( t^{(l)}_{d,tt}-t^{(l)}_{d,ts}\right)
  \;\;,\;\;
  t_{\text{Wa}}^{(l)}:=\half\left( t^{(l)}_{d,tt}+t^{(l)}_{d,ts}\right)\;\;,
\end{equation}
decouples thus the Faddeev equations of the doublet
channel~\cite{3stooges_doublet}:
\begin{eqnarray}
  \label{eq:faddeevWigner}
  {t_{\text{Ws}}^{(l)}\choose t_{\text{Wa}}^{(l)}}(p)&=&4\pi\;
  \calK^{(l)}(0;0,p)\;{1\choose -\half}\\
  &&+\frac{4}{\sqrt{3}\pi}\int\limits_0^\infty\deint{}{q}
  q^2\;\calK^{(l)}(0;q,p) 
    \begin{pmatrix}2&0\\
      0&-1\end{pmatrix} \frac{1}{q} \; {t_{\text{Ws}}^{(l)}\choose
      t_{\text{Wa}}^{(l)}}(q)\;\;. \non
\end{eqnarray}
$t_{\text{Wa}}^{(l)}$ obeys in this limit the same integral equation as the
quartet-channels, and $t_{\text{Ws}}^{(l)}$ is identical to the one for three
spin-less bosons~\cite{danilov,3stooges_doublet}. The problem to construct the
UV-behaviour of the three-body system with large two-body scattering length
simplifies hence to constructing the solution of just one integral equation:
\begin{equation}
  \label{eq:master}
  t^{(l)}_\lambda(p)=
  8\pi\lambda\;\lim\limits_{k\to0}\calK^{(l)}(\frac{3k^2}{4}-\gamma^2;k,p)
  +\frac{8\lambda}{\sqrt{3}\pi}\;\left(-1\right)^l\int
  \limits_0^\infty\frac{\deint{}{q}}{p}\;
  Q_l\left(\frac{p}{q}+\frac{q}{p}\right) \;t^{(l)}_\lambda(q)\;\;,
\end{equation}
where the amplitude $t^{(l)}_\lambda(p)$ depends only on the off-shell
momentum $p$, the partial wave $l$, and the spin-isospin combination
$\lambda$. The normalisation of the inhomogeneous part only provides the
overall scale of the solution and is hence irrelevant for the following.

In a slight abuse of terminology, the names ``deuteron'' and ``nucleon'' are
used in the remainder also when three identical bosons are considered.

\subsection{Divergences and Three-Body Forces at Higher Order}
\label{sec:higherorders}

Na\"ive dimensional analysis is based on the UV-behaviour of the scattering
amplitude: As outlined in the Introduction, a three-nucleon force is needed at
some order as counter-term to absorb cut-off dependence induced in the
physical amplitudes by divergences, as the ingredients of the Faddeev equation
are refined to include higher-order effects. The running of this
three-body force with the cut-off is then assumed to be of the same size as
its initial condition, which in turn must be determined from a three-body
datum. Equivalently, a three-body datum is needed at the same order as the
first divergence which must be absorbed by a three-nucleon interaction.  We
therefore discuss now the superficial degree of divergence of higher-order
corrections steming form the ''two-body sector'' of the theory.

\begin{figure}[!ht]
\begin{center}
  \includegraphics*[width=0.7\textwidth]{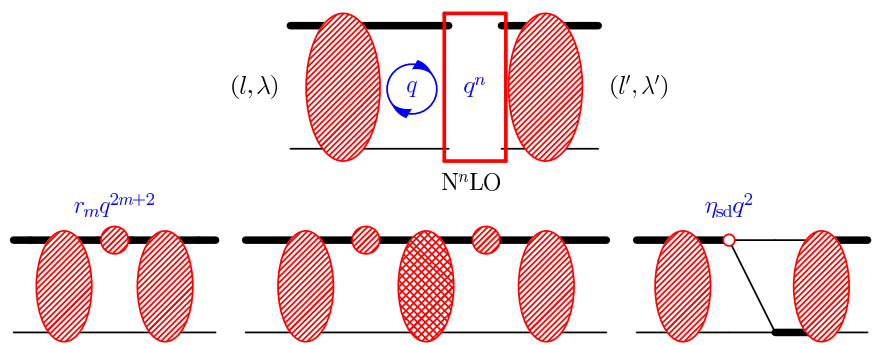}
\caption{Top: generic loop correction (rectangle) to the $Nd$-scattering
  amplitude at \NXLO{n}, proportional to $q^n$. Bottom, left to right:
  Exemplary higher-order contributions to $Nd$ scattering from the
  effective-range expansion (blob) and $\mathrm{SD}$-mixing (circle). Hatched
  ellipse: full off-shell amplitude. Notice that the external legs are
  on-shell.}
\label{fig:higherorders}
\end{center}
\end{figure}

As will be shown in the next Sub-section, the half off-shell amplitude at
large off-shell momenta $p\gg k$ is asymptotically given by
\begin{equation}
  \label{eq:solution}
  t^{(l)}_\lambda(p)\propto k^l\;p^{-s_l(\lambda)-1}\;\;,
\end{equation}
where $s_l(\lambda)$ is in general a complex number which depends on the
partial wave $l$ and channel $\lambda$. Higher-order corrections to
three-nucleon scattering can be obtained by perturbing around the LO solution,
see Fig.~\ref{fig:higherorders}. This is numerically
tricky~\cite{doubletNLO,4stooges} also because from next-to-next-to-leading
order (\NXLO{2}) on, the full LO-off-shell amplitude must be computed and
convoluted numerically with the corrections, see the centre bottom graph in
Fig.~\ref{fig:higherorders}. However, it allows a simple determination of the
order at which the first divergence occurs.

The asymptotic form of the amplitude at higher orders, and thus its
superficial degree of divergence, follows from a simple power-counting
argument: With $q$ the loop-momentum, the non-relativistic nucleon-propagator
scales asymptotically as $1/q^2$, its non-relativistic kinetic energy as
$q^2/M$, and a loop integral counts as $q^5/M$. The deuteron propagator
\eqref{eq:dprop} approaches $1/q$. Only corrections at \NXLO{n} which are
proportional to positive powers of $q$ are relevant in the UV-limit -- all
other corrections do not modify the leading-order asymptotics. They appear
together with some coefficients $C$ 
which encode short-distance phenomena and whose magnitude
is hence set by the breakdown-scale of the theory. 
Such corrections scale asymptotically as $(q/\LambdaNoPion)^n\sim Q^n$ for
dimensional reasons. The asymptotics of the generic \NXLO{n}-correction to the
LO amplitude represented by the rectangle in the top graph of
Fig.~\ref{fig:higherorders} is thus proportional to
\begin{equation}
  \label{eq:asymptotics}
  k^l\;q^{-(s_l(\lambda)+1)}\;\times\;
  \frac{q^5}{M}\;\frac{M}{q^2}\;\frac{1}{q}
  \;\left(\frac{q}{\LambdaNoPion}\right)^n\;\times\;
  k^{l^\prime}\;q^{-(s_{l^\prime}(\lambda^\prime)+1)}
  \propto k^l\;k^{l^\prime}\;
  q^{n-s_l(\lambda)-s_{l^\prime}(\lambda^\prime)}\;\;.
\end{equation}
We therefore identify $n-s_l(\lambda)-s_{l^\prime}(\lambda^\prime)$ as the
\emph{superficial degree of divergence} of a diagram. A correction at \NXLO{n}
diverges when
\begin{equation}
  \label{eq:twobodydivs}
  \Re[n-s_l(\lambda)-s_{l^\prime}(\lambda^\prime)]\geq0\;\;.
\end{equation}
While $s_l(\lambda)$ will turn out to be generically complex, this condition
depends only on its real part. The power $n$ of the higher-order insertion is
on the other hand a positive integer. Notice also that this formula is not
limited to scattering of three nucleons -- it applies equally well when
external currents couple to nucleons inside the box of
Fig.~\ref{fig:higherorders}.

\absatz Usually, higher-order corrections mix different partial waves, and
also Wigner-symmetric and Wigner-antisymmetric amplitudes in the
doublet-channel. They come from any combination of the following effects, some
of which are depicted in the lower panel of Fig.~\ref{fig:higherorders}:

\begin{enumerate}
\item Effective-range corrections to the deuteron propagator,
  \begin{equation}
    \label{eq:drefine}
    D(q_0,\qv)\to
    \frac{1}{\gamma-\sqrt{\frac{\qv^2}{4}-Mq_0}}\left[
    \sum\limits_{m=0}^\infty\frac{\frac{r_{0}}{2}\;(Mq_0-\frac{\qv^2}{4})+
    \sum\limits_{n=1}^\infty r_{n}\;(Mq_0-\frac{\qv^2}{4})^{n+1}}
    {\gamma-\sqrt{\frac{\qv^2}{4}-Mq_0}}
    \right]^m\;\;.
  \end{equation}
  With the coefficients $r_n\sim1/\LambdaNoPion^{2n+1}$ of natural size, these
  contributions are ordered by powers of $Q\sim q\sim\sqrt{Mq_0}$: The
  effective range $r_0$ enters at NLO as one insertion into the
  scattering-amplitude; $r_0^n$ at \NXLO{n} as $n$ insertions, etc. The
  correction $r_n$ starts contributing with one insertion at \NXLO{2n+1}. They
  modify the UV-limit of $D$ from $1/q$ in \eqref{eq:dforWigner} at \NXLO{n}
  to $q^{n-1}$.
  
\item As two-nucleon forces are non-central, different two-nucleon partial
  waves mix, e.g.~the \threeS- and \wave{3}{D}{1}-waves. This leads to mixing
  and spitting of partial waves also in the three-body problem, so that in
  general $s_l(\lambda)\not=s_{l^\prime}(\lambda^\prime)$. By
  parity-conservation, the lowest-order mixing appears for $l^\prime=l\pm2$.
  
\item The local vertex of two nucleons scattering via higher partial waves
  $L>0$ contains $2L$ positive powers of $q$ and mixes partial waves as well.
  
\item Insertions into $D$ which correct for the explicit breaking of
  Wigner-$SU(4)$-symmetry are proportional to the differences in the
  effective-range coefficients of the \oneS- and \threeS-system, e.g.~to
  $\gamma_s-\gamma_t$ or $q^2(\rho_{0,s}-\rho_{0,t})$, see
  also~\cite{4stooges}. This leads to mixtures with $l=l^\prime$ but
  $\lambda\not=\lambda^\prime$.
  
\item Other corrections of less importance, like relativistic corrections to
  the deuteron propagator $D$ and to the nucleon propagator.
\end{enumerate}

\begin{figure}[!ht]
\begin{center}
  \includegraphics*[width=0.7\textwidth]{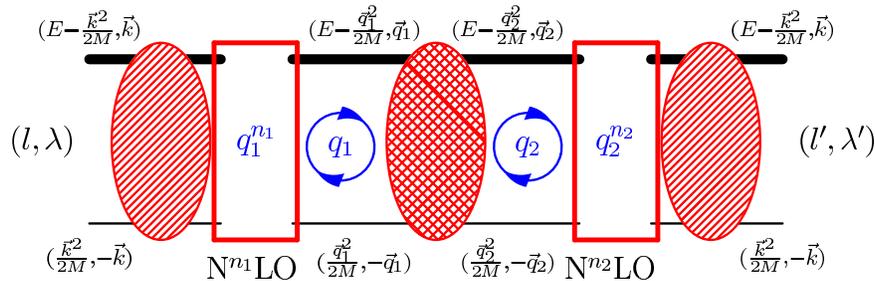}
\caption{Exemplary graph containing the full off-shell amplitude. Its
  kinematics is defined after the integrations over the energy-variables
  $q_{1,0}$ and $q_{2,0}$ are performed.}
\label{fig:fulloffshell}
\end{center}
\end{figure}

A longer remark is appropriate for corrections in which the LO full off-shell
amplitude $t^{(l)}_\lambda$ is sandwiched between two loop-momenta $q_1,q_2$.
They lead to overlapping divergences in these two variables. Examples are
given in the centre bottom graph of Fig.~\ref{fig:higherorders}, or for a
correction at \NXLO{n_1+n_1} involving one full off-shell amplitude in
Fig.~\ref{fig:fulloffshell}.
Its asymptotics is for each off-shell momentum determined by the same Faddeev
equation \eqref{eq:faddeev} with the only difference that the cm-energy $E$
and -momentum $k$ become independent variables, with the on-shell point at
$k=p=\sqrt{4(ME+\gamma^2)/3}$. In the asymptotic region $E\ll k,p$, the
integral equations for both off-shell momenta in the kinematics defined in
Fig.~\ref{fig:fulloffshell} are hence in analogy to \eqref{eq:master} given
by:
\begin{equation}
  \label{eq:offshell}
  \begin{split}
  t^{(l)}_\lambda(k,p)&=\dis
  8\pi\lambda\;\frac{(-1)^l}{kp}\;Q_l\left(\frac{p}{k}+\frac{k}{p}\right)
  +\frac{8\lambda}{\sqrt{3}\pi}\;\left(-1\right)^l\int
  \limits_0^\infty\frac{\deint{}{q}}{p}\;
  Q_l\left(\frac{p}{q}+\frac{q}{p}\right) t^{(l)}_\lambda(k,q)\\
  &=\dis
  8\pi\lambda\;\frac{(-1)^l}{kp}\;Q_l\left(\frac{p}{k}+\frac{k}{p}\right)
  +\frac{8\lambda}{\sqrt{3}\pi}\;\left(-1\right)^l\int
  \limits_0^\infty\frac{\deint{}{q}}{k}\;
  Q_l\left(\frac{k}{q}+\frac{q}{k}\right) t^{(l)}_\lambda(q,p)
  \end{split}
\end{equation}
It is symmetric under the interchange of $k$ and $p$. In analogy to the
solution of the half off-shell amplitude in the next Sub-section, its
asymptotics is constructed in App.~\ref{app:offshell} as:
\begin{equation}
  \label{eq:offshellsol}
  t^{(l)}_\lambda(k,p)\propto\left\{
   \begin{array}{l}
     \dis\frac{1}{kp}\left(\frac{k}{p}\right)^{s_l(\lambda)}\mbox{ for $p>k$}
     \\[2ex]
     \dis\frac{1}{kp}\left(\frac{p}{k}\right)^{s_l(\lambda)}\mbox{ for $p<k$}
   \end{array}\right\}\propto\frac{1}{kp}
\end{equation}
The latter follows by analytic continuation to $p\sim k\gg E,\gamma$. One may
motivate this result by observing that in the absence of any other scale, this
is the result with both the correct mass-dimensions and symmetry-properties.
It is independent of the angular momentum and spin-parameter. The superficial
degree of divergence of Fig.~\ref{fig:fulloffshell} is now easily determined
for $q\sim q_1\sim q_2$ scaling alike:
\begin{eqnarray}
  && k^l\;q_1^{-(s_l(\lambda)+1)}\;\;\frac{q_1^5}{M}\;\frac{M}{q_1^2}
  \;\frac{1}{q_1}\;\times
  \;\left(\frac{q_1}{\LambdaNoPion}\right)^{n_1}\;\times
  \;\frac{1}{q_1q_2}\;\times
  \;\frac{q_2^5}{M}\;\frac{M}{q_2^2}\;\frac{1}{q_2}
  \;\left(\frac{q_2}{\LambdaNoPion}\right)^{n_2}\;\times\;
  k^{l^\prime}\;q_2^{-(s_{l^\prime}(\lambda^\prime)+1)}\non\\[1ex]&&
  \sim k^l\;k^{l^\prime}\;q^{n_1+n_2-s_l(\lambda)-s_{l^\prime}(\lambda^\prime)}
\end{eqnarray}  
The overlapping divergence $E,k\ll q_1,q_2$ is hence also included in the
previous estimate \eqref{eq:twobodydivs} when $n=n_1+n_2$. One readily
generalises to any number $j$ of insertions of full off-shell amplitudes with
two-nucleon interactions at \NXLO{n_i}, $i=1,\dots,j+1$, between them and the
initial and final half off-shell amplitudes. They all are covered by
\eqref{eq:twobodydivs}, with the higher-order correction at \NXLO{n},
$n=\sum_{i=1}^{j+1} n_i$.

\absatz None of the corrections listed above leads at a given order \NXLO{n}
to stronger modifications of the short-distance asymptotics than those induced
by effective-range corrections entering at the same order, and every order
contains also contributions from effective-range corrections.

\subsection{Short-Distance Asymptotics of the Amplitude}
\label{sec:amplitude}

We now just have to determine the unphysical short-distance behaviour of the
amplitude to infer from \eqref{eq:twobodydivs} at which order the first
three-body forces are needed to absorb divergences. Na\"ively,
$t_\lambda^{(l)}(p)$ should have the same asymptotics as each of the
individual diagrams which need to be summed at LO, see top row of
Fig.~\ref{fig:kinematics}. That means, the asymptotic form should be given by
the inhomogeneous or driving term as
\begin{equation}
  \label{eq:naive}
  t^{(l)}_\lambda(p)\propto\lim\limits_{k\to0}\calK^{(l)}(\frac{3k^2}{4};k,p)
  \propto\frac{k^l}{p^{l+2}}\;\;\mbox{ , i.e. }\;\;
  s_{l,\text{simplistic}}(\lambda)=l+1\;\;. 
\end{equation}
This ``simplistic'' application of a na\"ive dimensional estimate reflects the
expectation that three-body forces should enter only at high orders, and that
the asymptotics in higher partial waves should be suppressed by a centrifugal
barrier. Indeed, this estimate would lead from (\ref{eq:twobodydivs}) to the
finding that the three-body force in the $l$th partial wave -- containing at
least $2l$ derivatives -- occurs only at \NXLO{2l+2} and is in particular
independent of the spin-parameter $\lambda$. However, the three-body problem
consists already at leading order of an infinite number of graphs, see
Fig.~\ref{fig:kinematics}. As is well-explored for $\mathrm{S}$-waves, this
modifies the solution drastically. 

The integral equation \eqref{eq:master} can be solved exactly by a Mellin
transformation since its homogeneous term is scale-invariant and
inversion-symmetric; see Appendix~\ref{app:appendix} for details. An implicit,
transcendental, algebraic equation determines the asymptotic exponent
$s_l(\lambda)$:
\begin{equation}
  \label{eq:s}
  1=\left(-1\right)^l\;\frac{2^{1-l}\lambda}{\sqrt{3\pi}}\;
  \frac{\Gamma\left[\frac{l+s+1}{2}\right]\Gamma\left[\frac{l-s+1}{2}\right]}
  {\Gamma\left[\frac{2l+3}{2}\right]}\;
  {}_2F_1\left[\frac{l+s+1}{2},\frac{l-s+1}{2};
    \frac{2l+3}{2};\frac{1}{4}\right]\;\;.
\end{equation}  
It depends only on $\lambda$ and $l$. The function ${}_2F_1[a,b;c;x]$ is the
hyper-geometric series~\cite{Gradstein}. This formula comprises the main
mathematical result of this article, extends Danilov's result for
$l=0$~\cite{danilov}, and forms in particular the base to power-count all
three-body forces. However, not all of its solutions solve also the integral
equation: While both $s$ and $-s$ are together with their complex conjugates
solutions to the algebraic equation, only those amplitudes which converge for
$p\to\infty$ and for which the Mellin transformation exists are permitted.
Most notably, this constrains $\Re[s]>-1$, $\Re[s]\not=\Re[l]\pm2$; see
Appendix~\ref{app:appendix}. Furthermore, out of the infinitely many, in
general complex solutions for given $l$ and $\lambda$, only the one survives
as relevant in the UV-limit whose real part is closest to $-1$, i.e.~for which
$\Re[s_l(\lambda)+1]$ is minimal. We consider in the following only those
solutions which match these criteria.

\begin{figure}[!ht]
\begin{center}
  \includegraphics*[width=0.48\textwidth]{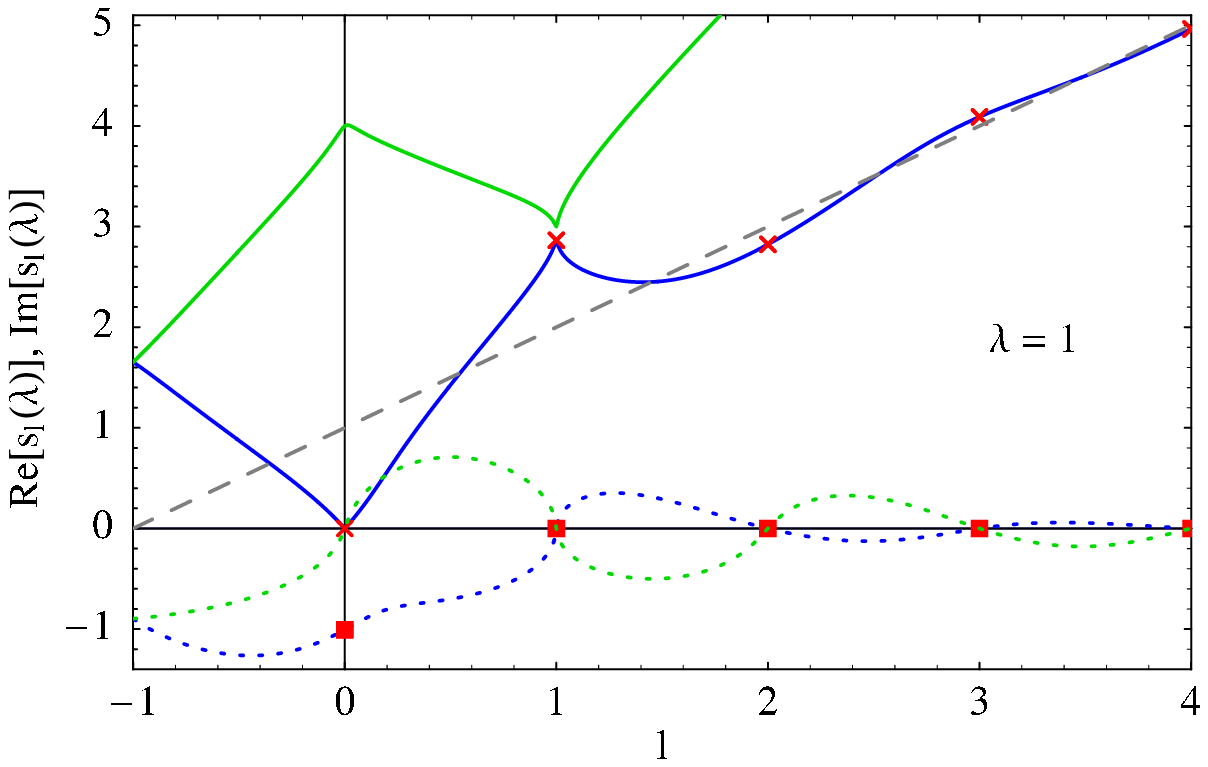}\hfill
  \includegraphics*[width=0.48\textwidth]{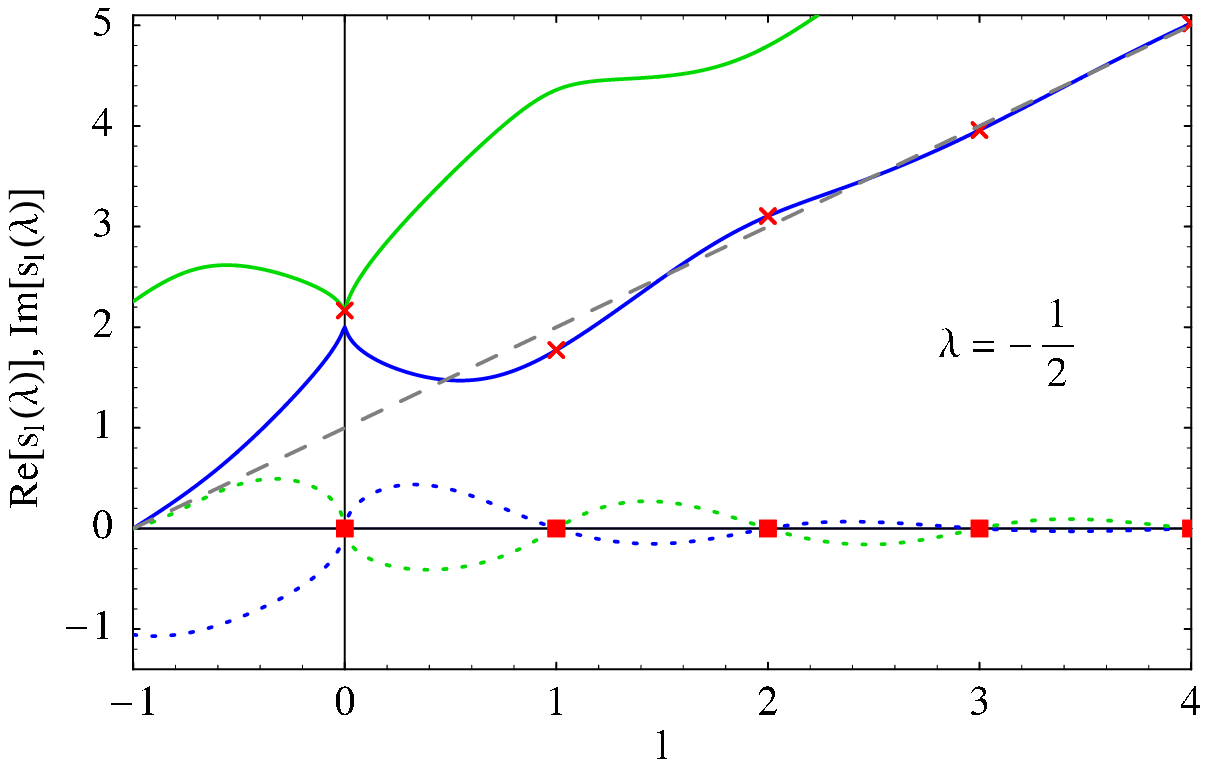}
\caption{The first two solutions $s_l(\lambda)$ at $\lambda=1$ (left) and
  $\lambda=-\half$. Solid (dotted): real (imaginary) part; dashed: simplistic
  estimate \eqref{eq:naive}; cross (square): real (imaginary) part of the
  asymptotics obtained by a fit of the full solution to the Faddeev equation
  \eqref{eq:faddeev} at large off-shell momenta. Dark/light: first/second
  solution. An Efimov effect occurs only for $|\Re[s]|<\Re[l+1]$, i.e. when
  the solid line lies below the dashed one, and $\Im[s]\not=0$.}
\label{fig:s-lambdafixed}
\end{center}
\end{figure}

\begin{figure}[!htb]
\begin{center}
  \includegraphics*[width=0.48\textwidth]{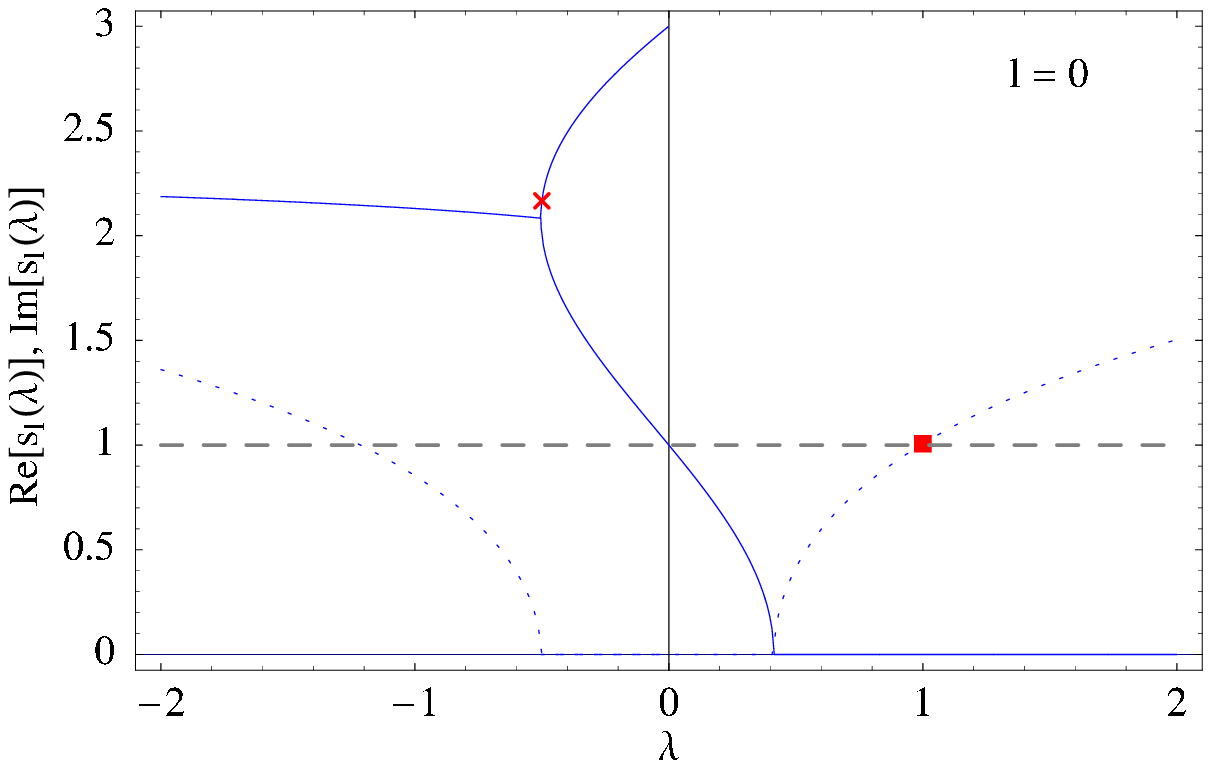}\hfill
  \includegraphics*[width=0.48\textwidth]{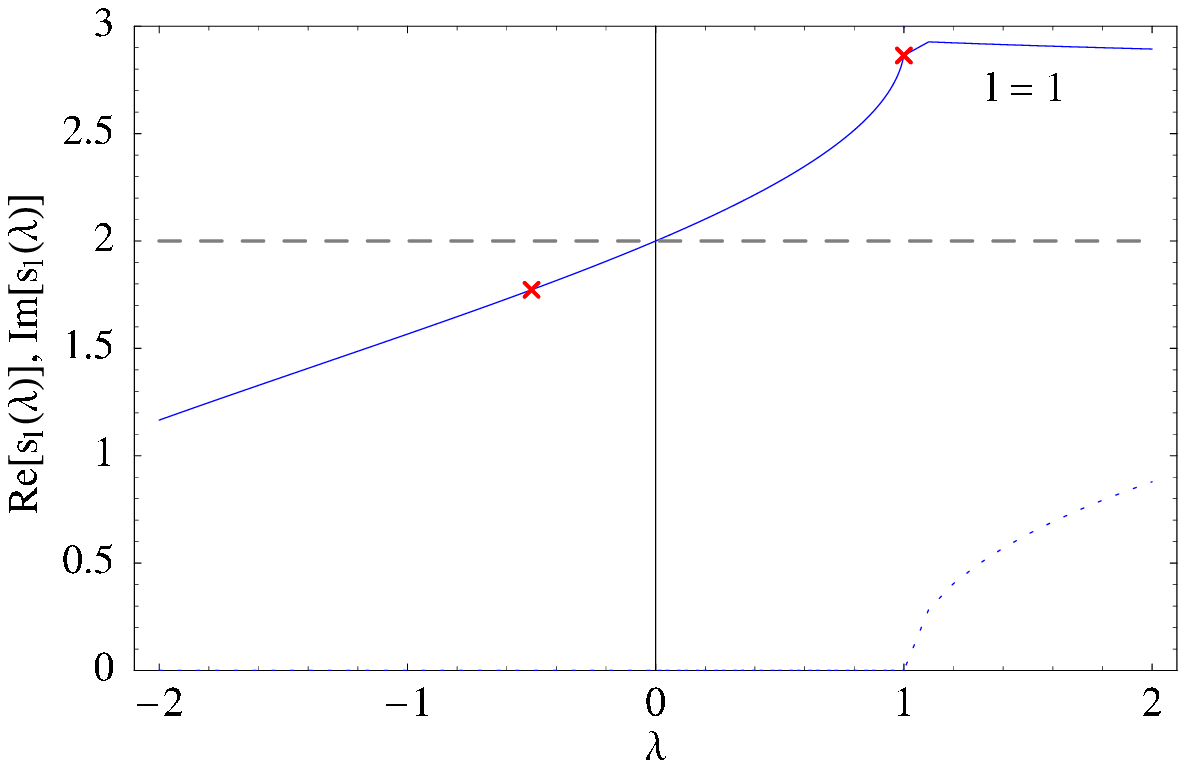}
  \\[2ex]
  \includegraphics*[width=0.48\textwidth]{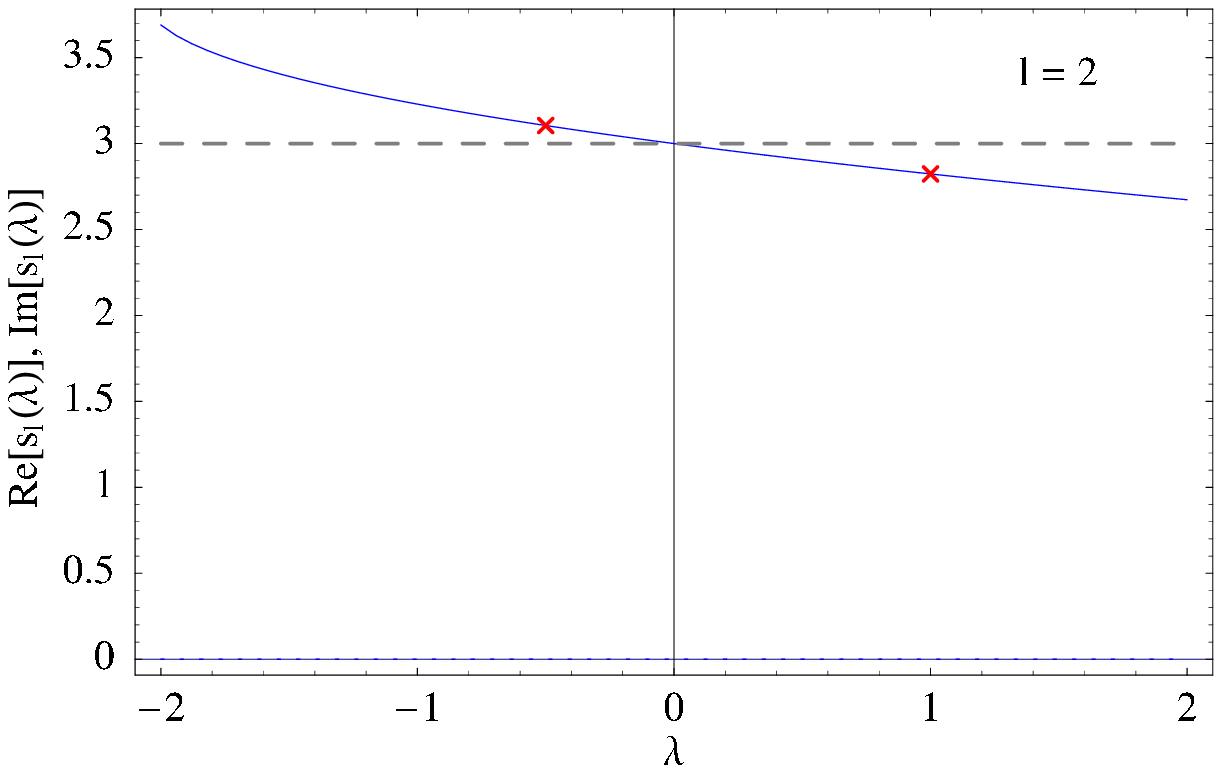}\hfill
  \includegraphics*[width=0.48\textwidth]{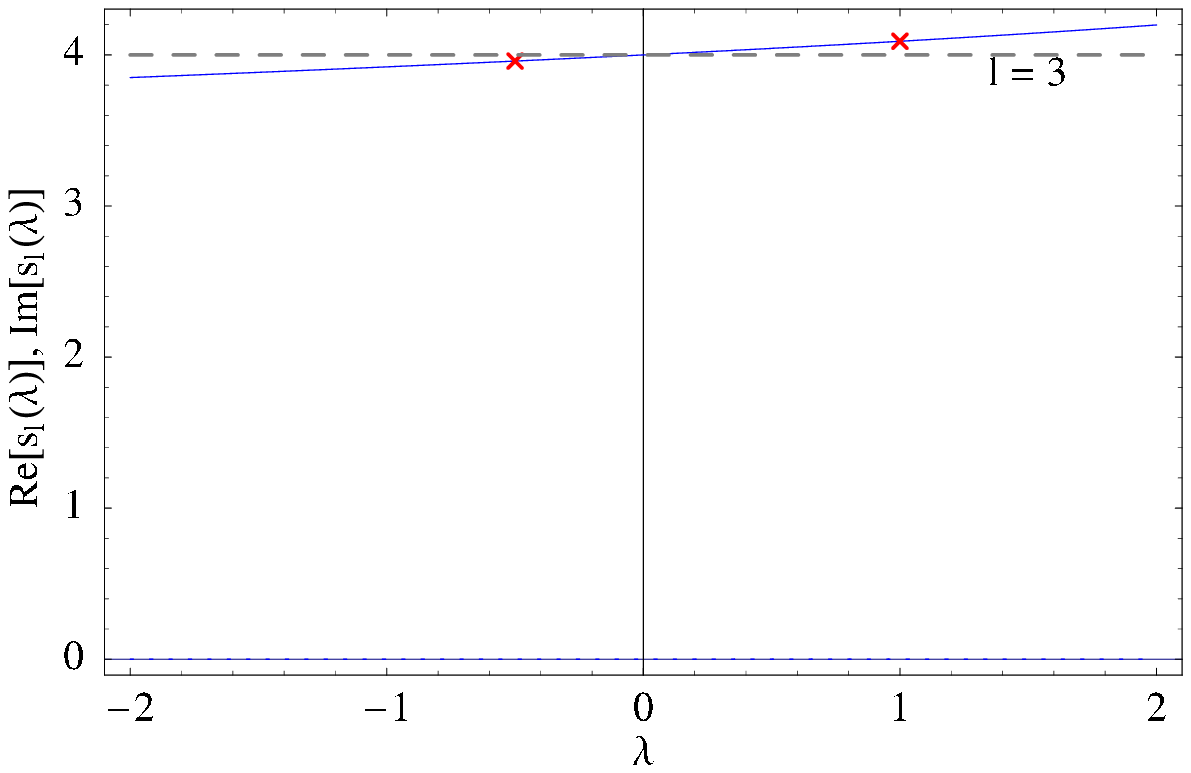}
\caption{$s_l(\lambda)$ at $l=\{0;1;2;3\}$. Notation as in
  Fig.~\ref{fig:s-lambdafixed}.}
\label{fig:s-lfixed}
\end{center}
\end{figure}

Plots of one of the values in the quadruplet of two-parameter functions $\{\pm
s_l(\lambda), \pm s_l^*(\lambda)\}$ at fixed $l$ and fixed $\lambda$,
respectively, are given in Figs.~\ref{fig:s-lambdafixed} and
\ref{fig:s-lfixed}. Table~\ref{tab:svalues} lists the first $s_l(\lambda)$ for
the partial waves $l\leq4$ and $\lambda=\{-\half;1\}$, compared to the
simplistic estimate \eqref{eq:naive}.

\begin{table}[!ht]
  \centering
  \begin{tabular}{|c||l|l||c|}
    \hline
   \rule[-1.5ex]{0ex}{4ex}
   partial wave $l$&$s_l(\lambda=1)$&$s_l(\lambda=-\half)$
   &$s_{l,\text{simplistic}}=l+1$\\
   \hline
   \hline
   \rule[-1.5ex]{0ex}{4ex}
   0&$1.00624\dots\;\ii$&2.16622\dots&1\\
   \hline
   \rule[-1.5ex]{0ex}{4ex}
   1&2.86380\dots&1.77272\dots&2\\
   \hline
   \rule[-1.5ex]{0ex}{4ex}
   2&2.82334\dots&3.10498\dots&3\\
   \hline
   \rule[-1.5ex]{0ex}{4ex}
   3&4.09040\dots&3.95931\dots&4\\
   \hline
   \rule[-1.5ex]{0ex}{4ex}
   4&4.96386\dots&5.01900\dots&5\\
   \hline
  \end{tabular}
  \caption{Solutions $s_l(\lambda)$ to \eqref{eq:s} for the most relevant
    physical systems.} 
  \label{tab:svalues}
\end{table}

\absatz Let us for the remainder of this sub-section investigate the rich
structure of this result. Branch points occur e.g.~for
$(l=0;\lambda\approx-\half)$, where imaginary parts open. Avoided crossings
are found e.g.~for $(l=1;\lambda=1)$, etc. While the solution is in general
complex, it is real for non-negative integer $l$ in the physical channels
discussed above, where $\lambda=\{1;-\half\}$. The only exception is the
imaginary solution for $(l=0;\lambda=1)$ first found by
Danilov~\cite{danilov}. It makes a three-body force in this channel mandatory
already at LO as the system would otherwise be unstable against collapse of
its wave-function to the origin, a phenomenon well-known to be related to the
Thomas- and Efimov-effects~\cite{thomas,efimovI} and giving rise to a
limit-cycle~\cite{3stooges_boson,wilson,Braaten:2003eu} manifesting itself in
the Phillips line~\cite{phillips,tkachenko,3stooges_doublet}. Its
interpretation is not the scope of this presentation; we only note that the
power-counting of three-body forces in this channel states that a new,
independent three-body force with $2l$ derivatives enters at
\NXLO{2l}~\cite{4stooges}. It must be seen as coincidence that the na\"ive
dimensional estimate in (\ref{eq:twobodydivs}) -- where $n>0$ was assumed
explicitly -- leads to the same conclusion.

In general, $s$ can be complex, as for example at $(l=0;\lambda<-\half)$,
$(l=1;\lambda>1)$ or $l$ non-integer, $\lambda=\{1;-\half\}$. In that case,
out of the four independent solutions, the ones with $\Re[s]\leq-1$ must be
eliminated as $t(p)$ does not converge for them. In cases like
$(l\in[-0.5819\dots;0.3446\dots];\lambda=1)$ where all four complex
solutions obey $\Re[s]>-1$, only the solutions with minimal
$\Re[s_l(\lambda)+1]$ survive, as shown above. These remaining two solutions
$s:=s_R\pm\ii s_I$ are equally strong and must be super-imposed:
\begin{equation}
  \label{eq:efimov}
  t^{(l)}_\lambda(p)\propto\frac{\sin[s_I\ln[p]+\delta]}{p^{s_R+1}}
\end{equation}
Usually, Fredholm's alternative forbids that both the homogeneous and
inhomogeneous integral equations have simultaneous solutions. Therefore, the
boundary conditions of the integral equation fix the phase $\delta$ to a
unique value. However, when the kernel of the Faddeev equation
\eqref{eq:faddeev} is singular, this operator has no inverse and Fredholm's
theorem does not apply. For this to occur, the solution to the integral
equation must be unique only up to a zero-mode of the homogeneous version,
i.e.
\begin{equation}
  \frac{8\lambda}{\sqrt{3}\pi}\;\left(-1\right)^l\;
  \int\limits_0^\infty\frac{\deint{}{q}}{p} \;
    Q_l\left(\frac{p}{q}+\frac{q}{p}\right) 
  a^{(l)}_{\lambda}(q)=a^{(l)}_{\lambda}(p)
\end{equation}
must have a non-trivial solution $a^{(l)}_{\lambda}(q)\not\equiv0$. As the
explicit construction in App.~\ref{app:integral} demonstrates, this is the
case if and only if $\Re[l+1]>|\Re[s]|$, i.e.~when $\Re[s]$ is smaller in
magnitude than the blind estimate $s_{l,\text{simplistic}}$, \eqref{eq:naive}.
In that case, a one-parameter family of solutions arises. A three-body force
is then necessary not to cure divergences but to absorb the dependence on the
free parameter $\delta$. Its initial condition is not constrained by two-body
physics but must be determined by a three-body datum. Thus, one finds a
limit-cycle for such systems, like for three spin-less bosons or the
Wigner-symmetric part of the \twoS-wave amplitude in $Nd$-scattering,
$(l=0;\lambda=1)$, discussed above. As easily read-up from
Figs.~\ref{fig:s-lambdafixed} and \ref{fig:s-lfixed}, this phenomenon occurs
for non-negative integer angular momentum only when $l=0$ and
$\lambda>3\sqrt{3}/(4\pi)$, where $\Re[s]=0$. However, an Efimov effect with
complex $s$ is often found for non-integer $l$, e.g.~for
$l\in[-0.3544\dots;0.5452\dots]$ in the three-boson case, $\lambda=1$. A
closer investigation will be interesting in view of a conjecture on
regularising the three-body system in Sect.~\ref{sec:conjectures}.

\begin{figure}[!htb]
\begin{center}
  \includegraphics*[width=0.48\textwidth]{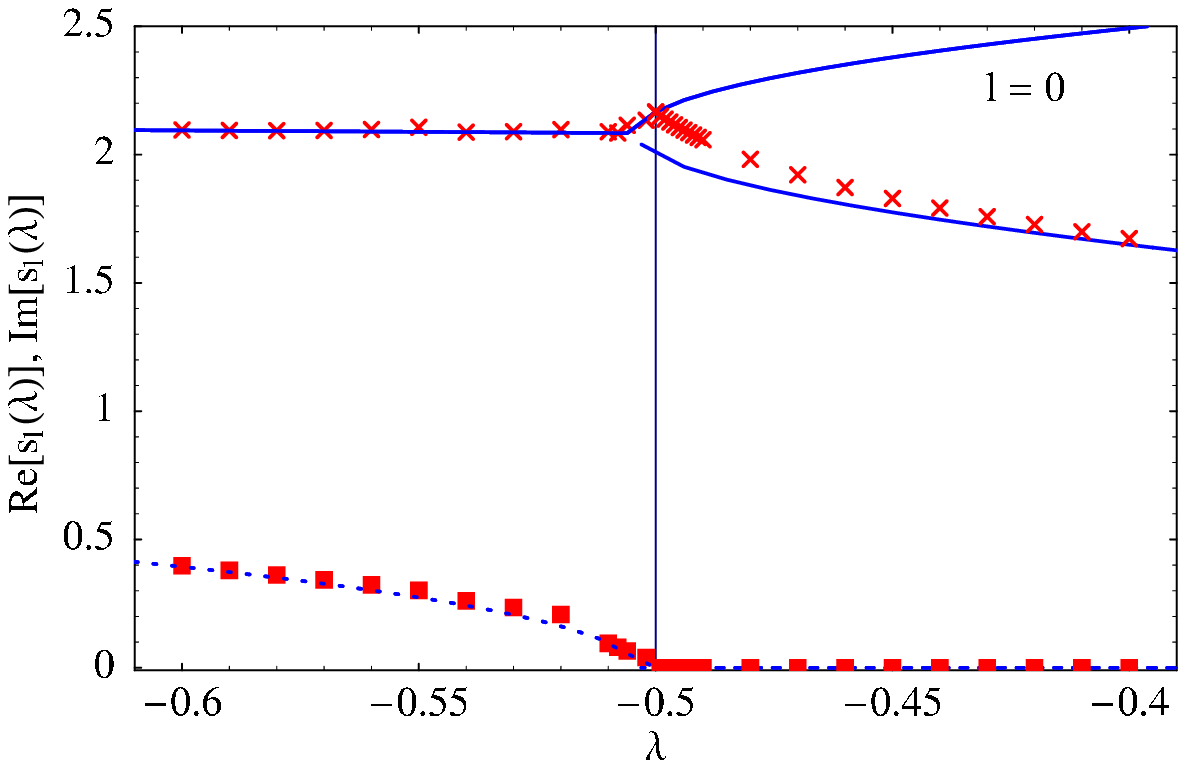} \hfill
  \includegraphics*[width=0.48\textwidth]{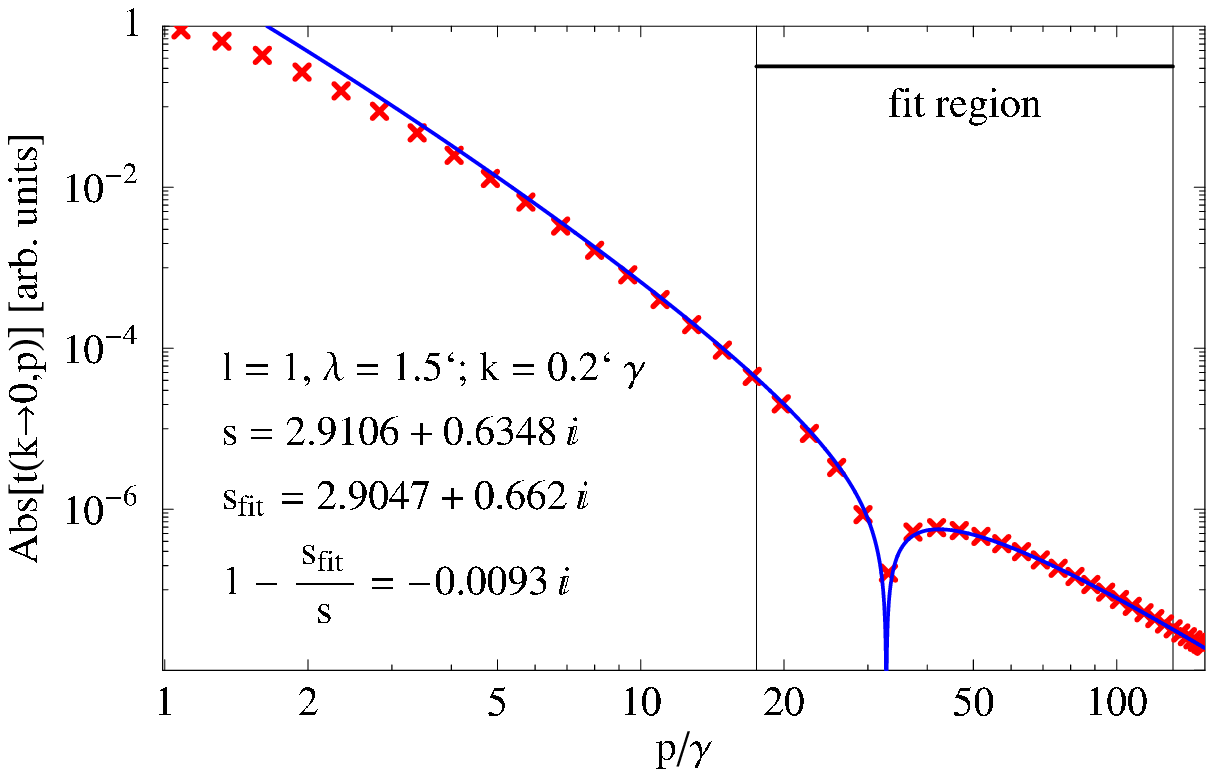}
\caption{Left: Numerical and analytical solution for $s_l(\lambda)$ at $l=0$
  around $\lambda=-\half$. Right: Numerical determination of $s_l(\lambda)$,
  exemplified for $l=1$, $\lambda=1.5$, comparing ``data'' (crosses) and the
  fitted function \eqref{eq:efimov} (solid line). Notation as in
  Fig.~\ref{fig:s-lambdafixed}. }
\label{fig:s-detail}
\end{center}
\end{figure}

A numerical investigation of the Faddeev equation \eqref{eq:faddeev} confirms
these findings. In order to compute a solution, one introduces a cut-off
$\Lambda$ which is un-physical and thus not to be confused with the
breakdown-scale $\LambdaNoPion$ of the EFT. The numerical values of
$s_l(\lambda)$ are found from fitting the half off-shell amplitude
$t^{(l)}_\lambda(E,k;p)$ at $E,k,\gamma\ll p\ll\Lambda$ to the asymptotic
forms \eqref{eq:solution} and \eqref{eq:efimov}. A grid of $100$ points is
easily enough for a numerical precision in $s$ of about $1\%$~\footnote{A
  simple Mathematica-code can be down-loaded from
  \texttt{http://www.physik.tu-muenchen.de/\~{}hgrie}.}. Agreement between
the numerical and analytical solution is excellent also at non-integer $l$ and
$\lambda\not=\{1,-\half\}$, see Fig.~\ref{fig:s-detail} besides
Figs.~\ref{fig:s-lambdafixed} and \ref{fig:s-lfixed} for examples.
Particularly interesting is in that context the neighbourhood around the
$\lambda=-\half$-solution in the $\mathrm{S}$-wave channel, $l=0$. Here, the
first solution to the algebraic equation \eqref{eq:s} is $s=2$, but the Mellin
transform does not exist at that point because $\Re[s]=l\pm2$. The system is
here also close to the branch-point at
$(\lambda=-0.50416\dots;s\approx2.0836\dots)$, where an imaginary part opens
for smaller $\lambda$. Another branch-point lies at
$(l=1;\lambda=1.0053\dots;s=2.93164\dots)$.

\absatz To summarise, the algebraic equation \eqref{eq:s} for $s_l(\lambda)$
provides asymptotic solutions of the form \eqref{eq:solution} to the
three-body Faddeev equation \eqref{eq:faddeev} for $\Re[s]>-1$ and
$\Re[s]\not=\Re[l\pm2]$. Only those solutions are relevant in the UV-limit
$p\gg\gamma,E,k$ for which $\Re[s+1]$ is minimal. The Efimov effect occurs
only if $\Im[s]\not=0$ and $|\Re[s]|<\Re[l+1]$, because only then is the
kernel of the integral equation not compact.

\subsection{Ordering Three-Body Forces}
\label{sec:ordering}

Although divergences can occur as soon as the superficial degree of divergence
$\Re[n-s_l(\lambda)-s_{l^\prime}(\lambda^\prime)]\geq0$, only those are
physically meaningful which can be absorbed by three-body counter-terms,
i.e.~by a local interaction between three nucleons in the given channels.
Na\"ive dimensional analysis does not construct the three-body forces. It thus
predicts some divergences which are absent when the diagram is actually
calculated. There is for example no Wigner-$SU(4)$ anti-symmetric three-body
force without derivatives~\cite{3stooges_doublet}, so that the divergence must
in this case be at least quadratic for infinite cut-off,
$\Re[n-s_l(\lambda)-s_{l^\prime}(\lambda^\prime)]\geq2$. For finite cut-off
$\Lambda$, a three-body force without derivatives can be constructed which is
non-local on a scale smaller than $1/\Lambda$ -- but appears of course local
at scales smaller than the break-down scale $\LambdaNoPion\lesssim\Lambda$ of
\EFTNoPion. As its coefficient dis-appears when the cut-off is sent to
infinity, it is in a renormalisation-group analysis classified as
``irrelevant''. Except for this example which is relevant below, this article
will however simply assume that the first three-body force enters with the
first divergence. To construct these forces in detail is left for future, more
thorough investigations.

\absatz The solution of \eqref{eq:s} approaches for large integer $l$ in the
physically most interesting cases $\lambda=\{1;-\half\}$ the simplistic
estimate of \eqref{eq:naive}: $s_\text{simplistic}= l+1$; see
Table~\ref{tab:lambda} and Figs.~\ref{fig:s-lambdafixed}, \ref{fig:s-lfixed}.
This reflects that the Faddeev equation should be saturated by the Born
approximation in the higher partial waves because of the ever-stronger
centrifugal barrier between the deuteron and the nucleon. Therefore,
three-body forces enter in most channels for all practical purposes at the
same order as suggested by the simplistic argument, namely \NXLO{l+l^\prime+2}
between the $l$th and $l^\prime$th partial waves. It is therefore convenient
to introduce the variable
\begin{equation}
  \label{eq:Deltadef}
  \Delta_l(\lambda):=s_l(\lambda)-(l+1)
\end{equation}
which parameterises how strongly simplistic and actual asymptotic form differ.
For $\Delta>0$, the superficial degree of divergence of the LO amplitude is
weaker than guessed by \eqref{eq:naive}.

\begin{table}[!ht]
  \centering
  \begin{tabular}{|cc|c||c|c||c|}
    \hline
   \multicolumn{3}{|c||}{channel}&  estimate &simplistic&\\[1ex]
   \cline{1-3}
   $(\lambda;l)$&$(\lambda^\prime;l^\prime)$
   &partial waves& $\Re[s_l(\lambda)+s_{l^\prime}(\lambda^\prime)]$&
   $l+l^\prime+2$&\\
   \hline
   \hline
   $(1;0)$&$(1;0)$&\wave{2}{S}{\text{Ws}}-\wave{2}{S}{\text{Ws}}&LO&
   \NXLO{2}&
   promoted\\
   $(1;0)$&$(-\half;0)$&\wave{2}{S}{\text{Ws}}-\wave{2}{S}{\text{Wa}}&
   \NXLO{2.2+2}&
   \multirow{2}{2cm}{\hspace*{\fill}\NXLO{2+2}\hspace*{\fill}}&\\
   $(-\half;0)$&$(-\half;0)$&
   \wave{2}{S}{\text{Wa}}-\wave{2}{S}{\text{Wa}}&\NXLO{4.3+2}&
   &demoted\\
   \hline
   $(1;0)$&$(-\half;2)$&\wave{2}{S}{\text{Ws}}-\wave{4}{D}{}&\NXLO{3.1}&
   \multirow{2}{2cm}{\hspace*{\fill}\NXLO{4}\hspace*{\fill}}&
   promoted\\
   $(-\half;0)$&$(-\half;2)$&\wave{2}{S}{\text{Wa}}-\wave{4}{D}{}&
   \NXLO{5.3}&&
   demoted\\
   \hline
   \hline
   $(1;1)$&$(1;1)$&\wave{2}{P}{\text{Ws}}-\wave{2}{P}{\text{Ws}}&
   \NXLO{5.7}&&
   demoted\\
   $(1;1)$&$(-\half;1)$&\wave{2}{P}{\text{Ws}}-\wave{2}{P}{\text{Wa}} ,
   \wave{2}{P}{\text{Ws}}-\wave{4}{P}{}
   &\NXLO{4.6}&\NXLO{4}&
   demoted\\
   $(-\half;1)$&$(-\half;1)$&
   \wave{2}{P}{\text{Wa}}-\wave{2}{P}{\text{Wa}} ,
   \wave{4}{P}{}-\wave{4}{P}{}&\NXLO{3.5}&&\\
   \hline
   \hline
   $(-\half;0)$&$(-\half;0)$&
   \wave{4}{S}{}-\wave{4}{S}{}&\NXLO{4.3+2}&\NXLO{2+2}&
   demoted\\
   $(-\half;0)$&$(1;2)$&\wave{4}{S}{}-\wave{2}{D}{\text{Ws}}&\NXLO{5.0}&
   \multirow{2}{2cm}{\hspace*{\fill}\NXLO{4}\hspace*{\fill}}&
   demoted\\
   $(-\half;0)$&$(-\half;2)$&\wave{4}{S}{}-\wave{2}{D}{\text{Wa}} ,
   \wave{4}{S}{}-\wave{4}{D}{}&
   \NXLO{5.3}&&
   demoted\\
   \hline
   \hline
   $(1;2)$&$(1;2)$&\wave{2}{D}{\text{Ws}}-\wave{2}{D}{\text{Ws}}&
   \NXLO{5.6}&&\\
   $(1;2)$&$(-\half;2)$&\wave{2}{D}{\text{Ws}}-\wave{2}{D}{\text{Wa}} ,
   \wave{2}{D}{\text{Ws}}-\wave{4}{D}{}
   &\NXLO{5.9}&\NXLO{6}&\\
   $(-\half;2)$&$(-\half;2)$&
   \wave{2}{D}{\text{Wa}}-\wave{2}{D}{\text{Wa}} ,
   \wave{4}{D}{}-\wave{4}{D}{}&\NXLO{6.2}&&\\
   \hline
  \end{tabular}
  \caption{Order $n_0$ at which the leading three-nucleon force enters for the
    lowest channels $l,l^\prime\leq2$, comparing the simplistic estimate
    \eqref{eq:naive} and the actual values
    (\ref{eq:twobodydivs}/\ref{eq:s}). The list follows the physical partial
    wave mixing, and the sub-script $\text{Ws}$ ($\text{Wa}$) denotes the
    Wigner-symmetric (anti-symmetric) contribution. In the
    \wave{2}{S}{\text{Wa}}- and \fourS-channels, the absence of a
    three-nucleon force without derivatives is taken into account by the
    factor ``$+2$''. The last column indicates whether the three-body force is
    stronger (``promoted'') or weaker (``demoted'') than the simplistic
    estimate suggests. When the difference between the two is in magnitude
    smaller than $0.5$, they are quite arbitrarily assumed to enter at the
    same order.} 
  \label{tab:ordering}
\end{table}

In the lower partial waves $l,l^\prime\leq2$, however, the blind expectation
deviates substantially from the exact solution; see Table \ref{tab:ordering}.
For $(l=0;\lambda=1)$, for example, $s_{l,\text{simplistic}}=l+1$
under-estimates the short-distance asymptotics of $t(p)$, while
$s_0(1)=1.006\dots\ii$ is even imaginary. A limit-cycle signals that one must
include a three-body force already at LO, as briefly hinted upon above.
Multiple insertions of three-body forces are not suppressed.
    
For two partial waves, the formula \eqref{eq:naive} substantially
\emph{over-estimates} the asymptotics of $t_\lambda^{(l)}$. Therefore, a
three-body force is in channels which involve these partial waves
\emph{weaker} than predicted by a simplistic application of na\"ive
dimensional analysis. Consider first the case of three bosons with
$(l=1;\lambda=1)$: $s=2.86\dots{}>s_\text{simplistic}=2$, $\Delta=0.86\dots$.
While the first divergence from the two-body sector arises in this partial
wave at \NXLO{5.72}, one would -- following \eqref{eq:naive} -- have predicted
the first three-body force as necessary already at \NXLO{4}. It is in this
channel hence demoted by $\approx1.7$ orders.

The situation is even more drastic in the \fourS-channel of $Nd$-scattering,
$(l=0;\lambda=-\half)$: Here, only divergences which are at least quadratic
are physical because the first three-body force must contain at least two
derivatives since the Pauli principle forbids a momentum-independent
three-nucleon force -- or equivalently, no Wigner-$SU(4)$ anti-symmetric
three-body force exists~\cite{3stooges_doublet}. Therefore, the divergence
condition \eqref{eq:twobodydivs} reads
$\Re[n-s_l(\lambda)-s_{l^\prime}(\lambda^\prime)\geq2$. Since
$s=2.16\dots{}>s_\text{simplistic}=1$, the first three-body force enters thus
not at \NXLO{4} but at least two orders higher, namely at \NXLO{6.33\dots}.

As as example for mixing between partial waves, consider the
\wave{4}{S}{\frac{3}{2}}-wave: It mixes with both the
\wave{4}{D}{\frac{3}{2}}-wave ($\lambda=-\half$) and the Wigner-symmetric and
anti-symmetric components of the \wave{2}{D}{\frac{3}{2}}-wave ($\lambda=1$ or
$-\half$). All of them are already close to the estimate
$s_{l,\text{simplistic}}=l+1=3>s_{l=0}(\lambda=-\half)$. Still, the first
divergences induced by this mixing start from \eqref{eq:twobodydivs} at
\NXLO{\approx5}, i.e.~approximately one order higher than blindly guessed.

More modifications induced by mixing and splitting of partial waves as well as
explicit breaking of the Wigner-$SU(4)$-symmetry are straight-forwardly
explored, but left to a future publication. Table \ref{tab:ordering}
summarises the findings for the physically most relevant three-body channels
$l,l^\prime\leq2$. Except in the \fourS-channel, it does not take into account
whether a three-body counter-term can actually be constructed at the order at
which the first divergence occurs. However, while this can make a three-body
force occur at a higher \emph{absolute} order than listed, the \emph{relative}
demotion of a three-body force to higher orders by modifications of the
superficial degree of divergence holds.

\absatz ``Fractional orders'' are a generic feature of \eqref{eq:s}, combined
with \eqref{eq:twobodydivs}. Consider again as example the case
$(l=0;\lambda=-\half)$, where the first two-body divergence appears formally
at $n=6.33\dots$, while including a two-body correction with fractional order
is of course impossible: Clearly, the \NXLO{7}-amplitude diverges without
three-body forces, but one could also argue that it is prudent to include a
three-body force already at \NXLO{6} because the higher-order correction to
the amplitude converges only weakly, namely as $q^{-0.33}$. Therefore, the
integral over $q$ becomes unusually sensitive to the amplitude at large
off-shell momenta, above the breakdown-scale $\LambdaNoPion$ of the EFT, and
therefore to details of Physics at distances on which the EFT is not any more
valid. In the more drastic case $s_{l=4}(\lambda=-\half)=5.02\dots$,
$\Delta=0.02\dots$, the first three-body force enters following
\eqref{eq:twobodydivs} at \NXLO{10.04}. Therefore, no divergence arises in the
two-body sector before \NXLO{11}, but it seems reasonable to include it
already at \NXLO{10} because the higher-order corrections from two-nucleon
insertions converge then generically to a cut-off independent result only very
slowly, namely as $q^{-0.04}$.

Na\"ive dimensional analysis cannot decide the question at which order
precisely a ``fractional divergence'' gives rise to a three-body force as it
argues on a diagram-by-diagram basis, missing possible cancellations between
different contributions at the same order. One way to settle it is to see
whether the cut-off dependence of observables follows the pattern required in
EFT. Recall that \NXLO{n} corrections contribute to observables typically as
\begin{equation}
  \label{eq:powercounting}
  Q^{n}=
  \left(\frac{p_\mathrm{typ}}{\LambdaNoPion}\right)^{n}
\end{equation}
compared to the LO result and that low-energy observables must be independent
of an arbitrary regulator $\Lambda$ up to the order of the expansion. In other
words, the physical scattering amplitude must be dominated by integrations
over off-shell momenta $q$ in the region in which the EFT is applicable,
$q\lesssim\LambdaNoPion$. As argued e.g.~by Lepage~\cite{Lepage:1997cs}, one
can therefore estimate sensitivity to short-distance Physics, and hence
provide a reasonable error-analysis, by employing a momentum cut-off $\Lambda$
in the solution of the Faddeev equation and varying it between the
breakdown-scale $\LambdaNoPion$ and $\infty$. If observables change over this
range by ``considerably'' more than $Q^{n+1}$, a counter-term of order $Q^n$
should be added. This method is frequently used to check the power-counting
and systematic errors in \EFTNoPion with three nucleons, see e.g.~most
recently~\cite{improve3body}. A similar argument was also developed in the
context of the EFT ``with pions'' of Nuclear
Physics~\cite{Bernard:2003rp,Epelbaum:2003gr}. Such reasoning goes however
beyond the clear prescription according to which only divergences make the
inclusion of counter-terms mandatory and opens the way to a softer criterion
-- which is obviously formulated rigorously only with great difficulty. How to
treat ``fractional orders'' in a well-prescribed and consistent way must thus
be investigated further.

\subsection{How Three-Body Forces Run}
\label{sec:analysing3bf}

\begin{figure}[!htb]
\begin{center}
  \includegraphics*[width=\textwidth]{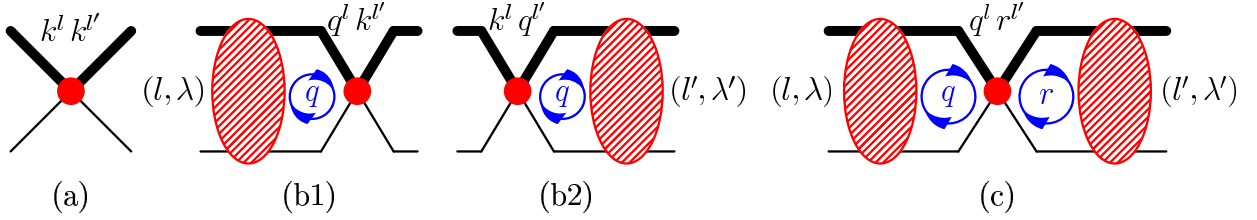}
\caption{Generic leading corrections from three-body forces which can be
  re-written as $Nd$-interactions with at least $l+l^\prime$ derivatives.}
\label{fig:3bodies}
\end{center}
\end{figure}

Before turning to practical consequences of these observations, let us for a
moment investigate how the strengthes of three-body forces have to scale with
$q$ in order to absorb the divergences \eqref{eq:asymptotics} from two-body
interactions. In contradistinction to the above considerations where the
specific form of the three-body force did not enter, we now limit the
discussion to those three-body forces which can be re-written as
deuteron-nucleon interactions. Clearly, all three-body forces which are needed
to absorb divergences from two-nucleon effective-range corrections
proportional to $r_n$ fall into that class~\footnote{But not necessarily
  three-body forces which contribute to the mixing and splitting of partial
  waves.}. As discussed in Sect.~\ref{sec:higherorders}, this is no severe
restriction because every divergence contains such a piece. The leading
contributions are given by the diagrams of Fig.~\ref{fig:3bodies}.
At even higher orders, two- and three-body corrections occur simultaneously in
one graph. Let the three-body force between deuteron and nucleon with relative
incoming momentum $p$ and outgoing momentum $p^\prime$ scale as
\begin{equation}
  \label{eq:3bftree}
  p^l\;p^{l^\prime}\;\;\frac{h}{\LambdaNoPion^{l+l^\prime+2}}
\end{equation}
where the dimension-less coupling $h$ encodes the short-distance details of
three interacting nucleons which are not resolved in \EFTNoPion. It absorbs
hence also the divergences generated at order $s_l(\lambda)+s_{l^\prime}$ from
two-body interactions, \eqref{eq:asymptotics}, to render the result at this
order insensitive of unphysical short-distance effects. The parameter $h$ must
thus formally scale as $q$ to some power $\alpha$ which is determined such
that at least one of the three-body force graphs appears at the same order as
the divergence. The graphs scale and diverge as:
\begin{equation}
  \begin{array}{lll}
  \label{eq:3bfscaling}
  \mbox{(a)}& \;\sim\; k^l\;k^{l^\prime}\;\;q^\alpha\;\;&\mbox{ , no
    divergence }
  \\[1ex]
  \mbox{(b1)}&\;\sim\; k^l\;k^{l^\prime}\;\;q^{\alpha-\Delta_l(\lambda)}\;\;
  &\mbox{ , diverges for }  \Re[\Delta_l(\lambda)]\leq0
  \\[1ex]
  \mbox{(b2)}&\;\sim\;
  k^l\;k^{l^\prime}\;\;q^{\alpha-\Delta_{l^\prime}(\lambda^\prime)}\;\; 
  &\mbox{ , diverges for }  \Re[\Delta_{l^\prime}(\lambda^\prime)]\leq0
  \\[1ex]
  \mbox{(c)} &\;\sim\; k^l\;k^{l^\prime}\;\;
  q^{\alpha-\Delta_l(\lambda)-\Delta_{l^\prime}(\lambda^\prime)}\;\;&\mbox{ ,
    diverges for }  \Re[\Delta_l(\lambda)]\leq0 \mbox{ or }
  \Re[\Delta_{l^\prime}(\lambda^\prime)]\leq0 
  \end{array}
\end{equation}
The tree-level contribution is of course free of divergences. Notice that the
three-body force $h$ absorbs for non-integer $s$ also the non-analytic piece
of the divergence \eqref{eq:asymptotics}, and in particular the phase $\delta$
when $\Im[s]\not=0$, see \eqref{eq:efimov}. This piece is non-analytic in the
un-physical off-shell momentum $q$, but of course analytic in the low-energy
momentum $k$. 

The graphs containing three-body forces enter at the same order for all
channels which follow the simplistic estimate of \eqref{eq:naive},
i.e.~$\Delta_l(\lambda),\Delta_{l^\prime}(\lambda^\prime)=0$. Then, all
three-body corrections (a-c) occur at the same order $\alpha$, the three-body
force counts as $h\sim q^\alpha=q^{l+l^\prime+2}$, and the logarithmic
divergences of the loop-diagrams (b1/2,c) are absorbed into $h$ as well. To
determine the order at which a three-body force enters, it is therefore
sufficient to count in this case its mass-dimension, which is also given by
$l+l^\prime+2$, see \eqref{eq:3bftree}. This is nearly realised in the higher
partial waves, where $|\Delta|$ is usually not bigger than $0.3$.

For $(l=0;\lambda=1)$, however, $\Re[\Delta]=-1$ and $\alpha=0$. Now, multiple
insertions of three-body forces are not suppressed and the Efimov effect
mandates including the three-body force in the LO Faddeev equation. The
strength $h$ depends on the arbitrary phase $\delta$, showing a
limit-cycle~\cite{3stooges_boson,wilson,Braaten:2003eu} as manifested in the
Phillips line~\cite{phillips,tkachenko,3stooges_doublet}. Interestingly, the
diagrams (a,b1/2) -- as well as their analogues with higher-order
three-nucleon forces -- now become following \eqref{eq:3bfscaling} formally
corrections of higher order, as found numerically in Ref.~\cite{doubletNLO}.

As seen in the previous Section, three-body forces appear in many channels at
higher orders than expected. This now also re-groups the graphs containing
three-body forces. According to the scaling properties \eqref{eq:3bfscaling},
the tree-level diagram (a) is in the $(l=1;\lambda=1)$-channel $\approx1.7$
orders weaker than the leading three-body diagram (c) because
$\Delta=0.86\dots$. It can therefore safely be neglected when absorbing the
leading divergences from two-body insertions. The graphs (b1/2) are down by
$\approx0.9\dots$ orders. For $(l=0;\lambda=-\half)$, this is even more
pronounced: With $\Delta=1.16\dots$, the tree-level three-body contribution
(a) is suppressed by more than two, and (b1/2) by more than one order against
the sandwiched three-body graph (c), so that both can be neglected when the
leading divergences are absorbed into (c) only. Notice that all three-body
corrections converge for $\Delta>0$.

\absatz Possible overlapping divergences complicate a similar analysis in the
case of three-body forces which cannot be re-written as $Nd$-interactions,
warranting further investigations which are however not central to this
presentation.

\section{Consequences}
\setcounter{equation}{0}
\label{sec:consequences}

The first goal of this publication has been reached:
Eq.~\eqref{eq:twobodydivs} is an explicit formula for the order at which the
first three-body force must be added to absorb divergences. It depends on the
partial wave $l$ and channel $\lambda$ via the exponent $s_l(\lambda)$ which
characterises the asymptotic form of the half off-shell scattering matrix
$t^{(l)}_\lambda(E,k;p)$ and is determined by \eqref{eq:s}. A simplistic
application of na\"ive dimensional analysis, \eqref{eq:naive}, provides a good
estimate for all partial waves $l\geq2$. However, it over-rates three-body
forces of the three-nucleon system for example in the \fourS- and
\wave{2}{P}{}-channels, while it under-estimates them e.g.~in the \twoS-wave;
see the summary in Table~\ref{tab:ordering}. In the case of three spin-less
bosons, the $\mathrm{P}$-wave three-body interaction is weaker, while the
$\mathrm{S}$-wave interaction is stronger than the simplistic argument
suggests. With these findings, the EFT of three spin-less bosons and the
pion-less version of EFT in the three-nucleon system, \EFTNoPion, are
self-consistent field theories which contain the least number of counter-terms
at each order to ensure renormalisability. Each three-body counter-term gives
rise to one subtraction-constant which must be determined by a three-body
datum. Let us explore in this Section some physically relevant results which
can be derived from these findings.

\subsection{Context}
\label{sec:context}

\begin{figure}[!htb]
\begin{center}
  \includegraphics*[width=0.82\textwidth]{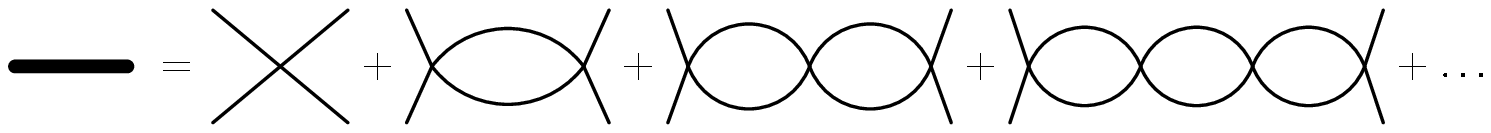}
\caption{Re-summation of the infinite number of LO two-body diagrams 
  into the deuteron propagator given by \eqref{eq:dprop}.}
\label{fig:twonucleonamp}
\end{center}
\end{figure}

\paragraph{Amending Na\"ive Dimensional Analysis:} As outlined in the
Introduction, power-counting by na\"ive dimensional analysis amounts for
perturbative theories to little more than counting the mass-dimensions of the
interactions~\cite{NDA}. In this case, only a finite number of diagrams
contributes at each order. When the LO amplitude is however non-perturbative,
i.e.~an infinite number of diagrams must be summed to produce shallow
bound-states, then the situation changes: The LO amplitude can follow for
large off-shell momenta a different power-law than the one which one obtains
when one considers the asymptotic form of each of the diagrams separately.
Then, the ``canonical'' application turns out to be too simplistic and must be
modified as in Sect.~\ref{sec:amplitude}. This is neither a failure of na\"ive
dimensional analysis, nor should it come completely unexpectedly. An example
is indeed already found in the two-body sector of \EFTNoPion. Recall that an
infinite number of two-body scattering diagrams are re-summed into the LO
deuteron propagator \eqref{eq:dprop} to produce the shallow two-body bound
state, Fig.~\ref{fig:twonucleonamp}.
Each of the diagrams diverges as $q^n$, with $n$ the number of loops. Their
sum \emph{converges} however as $1/q$ for large momenta,
see~\eqref{eq:dforWigner}. In this case, the solution is obtained by a
geometric series and the necessary changes are easily taken into account, see
the
reviews~\cite{bira_review,seattle_review,bedaque_bira_review,Braaten:2004rn}.
What may come as a surprise is that this can also happen when both all LO
diagrams separately and their re-summation are ultra-violet \emph{finite}. In
the three-body case, all diagrams actually show the \emph{same} power-law
behaviour $1/q^{l+1}$, see \eqref{eq:naive}. However, the re-summed form looks
very different, exhibiting a non-integer and even complex power-dependence
\eqref{eq:s}.
This does not occur for every system with shallow bound-states. For example,
the exchange of a Coulomb photon between two non-relativistic, charged
particles in non-relativistic QED scales asymptotically as $1/q^2$. The exact
solution of the Coulomb problem has the same scaling behaviour.

\paragraph{External currents:} The power-counting of three-body forces
developed above applies equally when external currents couple to the
three-nucleon system. The only change is that the higher-order interaction in
Fig.~\ref{fig:higherorders} becomes more involved, introducing also the
momentum- or energy-transfer from the external source as additional low-energy
scales.

\paragraph{Three-body forces at higher orders:} Another trivial extension is
to power-count three-body forces beyond the leading ones. In that case, the
superficial degree of divergence must by analyticity be larger than a positive
even integer, $\Re[n-s_l(\lambda)-s_{l^\prime}(\lambda^\prime)]\geq2m$,
$m\in2\mathbb{N}_0$. The higher-order three-body force contains $2m$
derivatives more than the leading one ($m=0$) and enters $2m$ orders higher.
We used this already to power-count the first three-body force in the
\fourS-channel. The power-counting based on na\"ive dimensional analysis
agrees for the \twoS-channel with the one which was recently established by a
more careful and explicit
construction~\cite{3stooges_doublet,doubletNLO,4stooges}.

\subsection{Conjectures}
\label{sec:conjectures}

\paragraph{Predicting the \fourS-scattering length:} The first conjecture
follows from the observation above that three-body forces are demoted in the
\fourS-wave from a \NXLO{4}-effect by two orders to \NXLO{6.3}. This has
immediate consequences for the quartet-$\mathrm{S}$ wave scattering-length of
the nucleon-deuteron system which has drawn substantial interest recently. Its
knowledge sets at present the experimental uncertainty in an indirect
determination of the doublet scattering length~\cite{boundcoherentexp}, which
in turn is well-known to be sensitive to three-body forces~\cite{phillips}. It
was determined repeatedly in \EFTNoPion at \NXLO{2}, with different methods to
compute higher-order corrections agreeing within the predicted
accuracy~\cite{2stooges_quartet,3stooges_quartet,pbhg,chickenpaper}, e.g.~most
recently~\cite{improve3body}:
\begin{equation}
  a(\fourS)=(
  \underbrace{\rule[-1ex]{0ex}{0ex}5.091}_\mathrm{LO}
  +\underbrace{\rule[-1ex]{0ex}{0ex}1.319}_\mathrm{NLO}
  -\underbrace{\rule[-1ex]{0ex}{0ex}0.056}_\text{\NXLO{2}})\;\fm=
  [6.35\pm0.02]\;\fm\;\;.
\end{equation}
The theoretical accuracy by neglecting higher-order terms is here estimated
conservatively by
$Q\sim\frac{\gamma\approx45\;\MeV}{\LambdaNoPion\approx\mpi}\approx\frac{1}{3}$
of the difference between the NLO- and \NXLO{2}-result. This agrees very well
with experiment~\cite{doublet_sca}, $[6.35\pm0.02]\;\fm$, albeit partial wave
mixing, iso-spin breaking and electro-magnetic effects are not present in
\EFTNoPion at \NXLO{2}. As the amplitude decays at large off-shell momenta as
$1/p^{3.16\dots}$, see Table~\ref{tab:lambda}, it is not surprising that
$a(\fourS)$ is to a very high degree sensitive only to the correct asymptotic
tail of the deuteron wave function. The first three-body force~\footnote{In
  the mixing between the \twoS-, \wave{2}{D}{}- and \wave{4}{D}{}-waves,
  three-body forces appear already at \NXLO{5}, see Table~\ref{tab:ordering},
  but they are irrelevant for the scattering length.}  enters not earlier than
\NXLO{6} -- taking a conservative approach as described above to round the
''fractional order''. Indeed, if the theoretical uncertainty decreases
steadily from order to order as it does from NLO to \NXLO{2}, then one should
be able to reach an accuracy of
$\pm(\frac{1}{3})^4\times0.06\;\fm\lesssim\pm0.001\;\fm$ with only two-nucleon
scattering data as input -- provided those in turn are known with sufficient
accuracy. Indeed, this is not much smaller than the range over which modern
high-precision potential-model calculations differ:
$[6.344\dots6.347]\;\fm$~\cite{Witala:2003mr,Friar:1999jd}. To use this number
hence as input into a determination of the doublet scattering-length as
$a(\twoS)=[0.645\pm0.003(\mathrm{exp})\pm
0.007(\mathrm{theor})]\;\fm$~\cite{boundcoherentexp} seems justified, and the
error induced by the theoretical uncertainty might actually be over-estimated.
Notice that if the three-nucleon force would occur in \EFTNoPion at \NXLO{4}
as the simplistic expectation \eqref{eq:naive} suggests, the error should be
of the order of $(\frac{1}{3})^2\times0.06\;\fm\approx0.007\;\fm$,
considerably larger than the spread in the potential-model predictions.

\paragraph{An alternative regularisation:} More speculative is the
possibility for a new regularisation scheme. In principle, \eqref{eq:s} gives
the asymptotics $s_l(\lambda)$ of the half off-shell amplitude for arbitrary
-- even complex -- $l$ and $\lambda$. As this function is largely analytic,
one could use analytic continuation for a ``partial wave regularisation'' of
the three-body system. This is particularly attractive to regulate the
limit-cycle problem of the \twoS-wave $(l=0;\lambda=1)$, whose practical
implications are at present mostly discussed by cut-off regularisation.
However, the algebraic solution to \eqref{eq:s} suffers -- as discussed in
Sect.~\ref{sec:amplitude} -- from constraints by branch-cuts and regions where
the Mellin transformation does not exist. In addition, a limit-cycle is
encountered only when $s$ has an imaginary part, and its real part is smaller
than the simplistic estimate \eqref{eq:naive}. However, the imaginary part
dis-appears in the vicinity of the physical point $(l=0;\lambda=1)$ only where
$s$ has a branch-cut, see Figs.~\ref{fig:s-lambdafixed} and
\ref{fig:s-lfixed}. Blankleider and Gegelia~\cite{Blankleider:2000vi}
attempted to use analyticity in $\lambda$ at fixed $l=0$ to regulate the
three-boson problem at LO without resorting to three-body forces to stabilise
the system.

\subsection{Caveats}
\label{sec:caveats}

Weak points in the derivation should also be summarised:

\begin{enumerate}
\item As always in na\"ive dimensional analysis, one obtains only the
  superficial degree of divergence. Except in the \fourS-channel, the
  classification of Table~\ref{tab:ordering} does not take into account
  whether a three-body counter-term can actually be constructed at the order
  at which the first divergence occurs. However, while the actual degree of
  divergence must usually be determined by an explicit calculation, it is is
  never larger than the superficial one. Therefore, a three-body force can
  possibly occur at a higher \emph{absolute} order than predicted by na\"ive
  dimensional analysis, but this applies then equally well to the simplistic
  estimate. Therefore, the \emph{relative} demotion of a three-body force to
  higher orders by the modified superficial degree of divergence holds.
  In this context, the three-body forces which can contribute in a given
  channel, and in particular to partial-wave mixing, should be constructed
  explicitly. Here, the symmetry principles invoked above can be helpful.
  
\item The divergence of each diagram was considered separately, missing
  possible cancellations between different contributions at a given order.
  This would again demote three-body forces to higher orders than determined
  by the superficial degree of divergence. It would also require a fine-tuning
  whose origin would have to be understood. When further re-summations of
  infinitely many diagrams should be necessary beyond LO, na\"ive dimensional
  analysis must be amended further. This could happen when the power-counting
  developed here does not accord to nuclear phenomenology.

\item Modifications by overlapping divergences should also be explored. Again,
  they weaken the degree of divergence, but are particularly important for
  those three-body forces which cannot be re-written as $Nd$-interactions.
  
\item The problem of ``fractional orders'': Since the LO amplitude involves an
  infinite number of graphs, equivalent to the solution of a Faddeev equation,
  the amplitude approaches for large half off-shell momenta generically a
  power-law behaviour $q^{-s_l(\lambda)-1}$ with irrational and even complex
  powers. Three-body forces contain therefore following
  Sect.~\ref{sec:analysing3bf} non-analytic pieces. We assume that this
  behaviour is changed at higher orders only by integer powers because
  higher-order corrections involve only a finite number of diagrams after the
  LO-graphs are summed into the Faddeev equation. At which concrete order a
  given three-nucleon interaction needs to be included to render observables
  cut-off independent can therefore become a question beyond the clear
  prescription according to which only divergences make the inclusion of
  counter-terms mandatory. A softer criterion is formulated rigorously only
  with great difficulty.
  
\item The two-nucleon propagator \eqref{eq:dprop} at the starting point of the
  derivation was taken to be already renormalised. This should pose no problem
  as the Faddeev equation was solved without further cut-offs, so that no
  overlapping divergences occur. Indeed, any ``scale-less'' regulator (like
  dimensional regularisation) will lead to the same result.
  
\item Partial wave-mixing and -splitting as well as mixing between
  Wigner-$SU(4)$ symmetric and anti-symmetric amplitudes should be considered
  in more detail. However, we saw already examples where these effects are
  suppressed. For example, the power-counting in the partial wave with the
  lowest angular-momentum amongst those that mix is unchanged, see the
  \fourS-\wave{4}{D}{\frac{3}{2}}-\wave{2}{D}{\frac{3}{2}}-mixing in
  Table~\ref{tab:ordering}. The higher the partial wave, the closer is its
  asymptotics to the simplistic expectation \eqref{eq:naive}.
  
\item We assumed -- as usual in EFT -- that the typical size of three-nucleon
  counter-terms is set by the size of their running. There are cases where the
  finite part of a counter-term is anomalously large and thus should be
  included already at lower orders than the na\"ive dimensional estimate
  suggests. One example is the anomalous iso-vector magnetic moment of the
  nucleon which is as large as the inverse expansion parameter of \EFTNoPion,
  $\kappa_1=2.35\approx1/Q$~\cite{Griesshammer:2000mi}. Such cases are however
  rare and must be justified with care. 
  
\end{enumerate}

Finally, Blankleider and Gegelia~\cite{Blankleider:2000vi} claimed in an
unpublished preprint 5 years ago that the \twoS-wave problem can be solved at
LO without resorting to a three-body force to stabilise the system against
collaps. According to them, if the Faddeev equation has multiple solutions,
then only one is equivalent to the series of diagrams drawn in
Fig.~\ref{fig:kinematics}. We focus in this article on the higher partial
waves where the Faddeev equation has -- as demonstrated in
Sect.~\ref{sec:amplitude} -- always unique solutions for integer $l>0$ and
$\lambda\in\{1;-\half\}$, so that the alleged discrepancy cannot arise.
Distracting the reader for a moment, one may however point out a few
observations which contradict the claim of Ref.~\cite{Blankleider:2000vi}. The
derivation of the Faddeev equation is just a special case of Schwinger-Dyson
equations, which are well-known to be derived in the path-integral formalism
without resort to perturbative methods, see
e.g.~\cite[Chap.~10]{ItzyksonZuber}. One resorts to a ``series of diagrams''
only for illustrative purposes like in Fig.~\ref{fig:kinematics}. Recall that
in the case of the three-body system, no small expansion parameter exists in
which this series can be made to converge absolutely.
In addition, and on a less formal level, well-known properties of the
three-body system like the Thomas and Efimov effects~\cite{thomas,efimovI} and
the Phillips line~\cite{phillips,tkachenko,3stooges_doublet} are not explained
under the assertions of Ref.~\cite{Blankleider:2000vi}. These universal
properties were recently also tested experimentally, e.g.~for particle-loss
rates in Bose-Einstein condensates near Feshbach resonances,
see~\cite{Braaten:2004rn} for a review.

\section{Conclusions and Outlook}
\setcounter{equation}{0}
\label{sec:conclusions}

In this article, the ordering of three-body contributions in three-body
systems also coupled to external currents was constructed systematically for
any EFT with only contact interactions and an anomalously large two-body
scattering length. Evading explicit calculations, the result is based on
na\"ive dimensional analysis~\cite{NDA}, improved by the observation that
because the problem is non-perturbative already at leading order, the solution
to the Faddeev integral equation does for large off-shell momenta not follow a
simplistic dimensional estimate, Sects.~\ref{sec:amplitude}
and~\ref{sec:context}. This was shown by constructing the analytical solution
to the Faddeev equation in that limit for arbitrary angular momentum and
spin-parameter. One could thus develop a ``partial-wave regularisation'' as an
alternative to regulate and renormalise the three-body system. A simplistic
approach to na\"ive dimensional analysis fails for systems which are
non-perturbative already at leading order.

In order to keep observables insensitive to the details of short-distance
Physics, one employs the canonical EFT tenet that a three-body force must be
included \emph{if and only if} it is needed as counter-term to cancel
divergences which can not be absorbed by renormalising two-nucleon
interactions. After determining the superficial degree of divergence of a
diagram which contains only two-nucleon interactions in
Sect.~\ref{sec:higherorders}, this was used in Sect.~\ref{sec:ordering} to
classify the relative importance of three-body interactions for each channel,
also for partial-wave mixing and splitting. With these results, the EFT of
three spin-less bosons and \EFTNoPion become self-consistent field theories
which contain the minimal number of parameters at each order to ensure
renormalisability and a manifest power-counting of all forces. Each such
three-body counter-term gives rise to one subtraction-constant which must be
determined by a three-body datum.

It must again be stressed that three-body forces are in \EFTNoPion added not
out of phenomenological needs. Rather, they cure the arbitrariness in the
short-distance behaviour of the two-body interactions which would otherwise
contaminate the on-shell amplitude, and hence make low-energy observables
cut-off independent on the level of accuracy of the EFT-calculation. Recall
that the theory becomes invalid at short distances as processes beyond the
range of validity of \EFTNoPion are resolved, namely the pion-dynamics and
quark-gluon sub-structure of QCD. Three-body forces are thus not introduced to
meet data but to guarantee that observables are insensitive to off-shell
effects. Only the combination of two-body off-shell and three-body effects is
physically meaningful.

Most of the three-nucleon forces in partial waves with angular momentum less
than $3$ have a \emph{weaker} strength than one would expect from a blind
application of na\"ive dimensional analysis, see Table~\ref{tab:ordering}.
This might seem an academic dis-advantage -- to include some higher-order
corrections which are not accompanied by new divergences does not improve the
accuracy of the calculation; one only appears to have worked harder than
necessary. However, it becomes a pivotal point when one hunts after three-body
forces in observables: In order to predict the experimental precision
necessary to dis-entangle these effects, the error-estimate of EFT is a
crucial tool. For many problems, this makes soon a major difference in the
question whether an experiment to determine three-body force effects is
feasible at all. One such consequence was discussed in
Sect.~\ref{sec:conjectures}: The \fourS-wave scattering length is fully
determined by two-nucleon scattering observables on the level of
$\pm0.001\;\fm$ according to the power-counting of \EFTNoPion developed here.
This is more than a factor of ten more accurate than the present experimental
number~\cite{doublet_sca}, but supported by the observation that all modern
high-precision two-nucleon potentials predict this observable to a similarly
high accuracy~\cite{Witala:2003mr,Friar:1999jd}. If a three-nucleon force
(even one saturated by pion-exchanges) would occur at the order at which it is
blindly expected, the spread in the potential-model predictions should be
considerably larger.

The consequences for other observables like the famed
$A_y$-problem~\cite{Huber:1998hu} should also be explored. This will be
particularly simple in \EFTNoPion because the theory is less involved and
Table~\ref{tab:ordering} sorts the three-nucleon forces according to their
strengthes, indicating also their symmetries and the channels in which they
contribute on the necessary level of accuracy. The conclusions, conjectures
and caveats of Sect.~\ref{sec:consequences} summarise a number of further
interesting directions for future research.

To classify the order at which a given two- or three-nucleon interaction
should be added in Chiral EFT, the EFT of Nuclear Physics with pions as
explicit degrees of freedom, I suggest to follow a path as for \EFTNoPion
which complements the so-far mostly pursued phenomenological approach: At
leading order, the theory must be non-perturbative to accommodate the
finely-tuned real and virtual two-body bound states in the $\mathrm{S}$-waves
of two-nucleon scattering. After that, only those local two- and three-nucleon
forces are added at each order which are necessary as counter-terms to cancel
divergences of the amplitudes at short distances. This mandates a more careful
look at the leading-order, non-perturbative scattering amplitudes to determine
their ultraviolet-behaviour and superficial degree of divergence, see
e.g.~\cite{Beane:2001bc} and references therein. It leads at each order and to
the prescribed level of accuracy to a cut-off independent theory with the
smallest number of experimental input-parameters. The power-counting is thus
not constructed by educated guesswork but by rigorous investigations of the
renormalisation-group properties of couplings and observables by EFT-methods.
Work in this direction is under way, see also~\cite{nogga}, and the future
will show its viability.


\section*{Acknowledgements}
I am particularly indebted to P.~F.~Bedaque and U.~van Kolck for discussions
and encouragement. H.-W.~Hammer provided valuable corrections to the
manuscript. The warm hospitality and financial support for stays at the INT in
Seattle and at the ECT* in Trento was instrumental for this research.  In
particular, I am grateful to the organisers and participants of the ``INT
Programme 03-3: Theories of Nuclear Forces and Nuclear Systems''. I also
acknowledge discussions with a referee concerning the equivalence of Faddeev
equations and series of diagrams, in which however no consensus was achieved --
as in previous exchanges on the same subject. This work was supported in part
by the Bundesministerium f\"ur Forschung und Technologie, and by the Deutsche
Forschungsgemeinschaft under contracts GR1887/2-2 and 3-1.

\newpage

\appendix
\section{Solving the Integral Equation}
\setcounter{equation}{0}
\label{app:appendix}

\subsection{Constructing the Solution}
\label{app:construction}

The Mellin transformation of a function $f(p)$, see e.g.~\cite{MorseFeshbach},
is defined as
\begin{equation}
  \label{eq:mellin}
  \calM[f;s]:=\int\limits_0^\infty\deint{}{p} p^{s-1}\;f(p)\;\;\;\;\;\;
  \mbox{ if }
  \int\limits_0^\infty\frac{\deint{}{p}}{p} |f(p)|^2 \mbox{ exists.}
\end{equation}
Applying this to both sides of (\ref{eq:master}) and using the faltung
theorem~\cite[Chap.~4.8]{MorseFeshbach}, one obtains the algebraic equation
\begin{equation}
  \calM[t;s]=8\pi\lambda\;
  \calM[\lim\limits_{k\to0}\calK^{(l)}(\frac{3k^2}{4}-\gamma^2;k,p);s]
  +\frac{8\lambda}{\sqrt{3}\pi}\;\left(-1\right)^l\;
  \calM[Q_l\left(x+\frac{1}{x}\right);s-1]\;\calM[t;s]\;\;,
\end{equation}
which is easily solved for $\calM[t;s]$. Thus, one now only has to apply an
inverse Mellin transformation,
\begin{equation}
  t(p)=\frac{1}{2\pi\ii}\;
  \int\limits_{c-\ii\infty}^{c+\ii\infty}\deint{}{s}p^{-s}\;\calM[t;s]\;\;,
\end{equation}
where the inversion contour must be placed in the strip $c$ in which all of
the original Mellin transformations exist.

However, there is no Mellin transform of $\calK^{(l)}$ in the limit
$\gamma,k\ll p$ because it is proportional to $(k)^l/p^{l+2}$. One therefore
has to resort for the inhomogeneous term to a slightly more complicated,
``half-plane'' transformation~\cite[Chap.~8.5]{MorseFeshbach}:
\begin{equation}
  \begin{split}
    \calM_-[\lim\limits_{k\to0}\calK^{(l)}(\frac{3k^2}{4}-\gamma^2;k,p);s]:=
    \int\limits_0^1\deint{}{p} p^{s-1}\;
    \lim\limits_{k\to0}\calK^{(l)}(\frac{3k^2}{4}-\gamma^2;k,p)
    \;\propto\;\frac{k^l}{s-l-2}
    \\
    \calM_+[\lim\limits_{k\to0}\calK^{(l)}(\frac{3k^2}{4}-\gamma^2;k,p);s]:=
    \int\limits_1^\infty\deint{}{p} p^{s-1}\;
    \lim\limits_{k\to0}\calK^{(l)}(\frac{3k^2}{4}-\gamma^2;k,p)
    \;\propto\;-\frac{k^l}{s-l-2}
  \end{split}
\end{equation}
These Mellin transforms $\calM_-$ and $\calM_+$ exist for $\Re[s]>\Re[l+2]$
and $\Re[s]<\Re[l+2]$, respectively. The solution to the integral equation is
in this case given by~\cite[eq.~(8.5.43)]{MorseFeshbach}:
\begin{eqnarray}
  \label{eq:aa}
  t_\lambda^{(l)}(p)=
  \frac{1}{2\pi\ii}\bigg[&&
  \int\limits_{\sigma_--\ii\infty}^{\sigma_-+\ii\infty}
  \deint{}{s}p^{-s}\;
  \frac{8\pi\lambda\;\calM_-[\lim\limits_{k\to0}\calK^{(l)}
    (\frac{3k^2}{4}-\gamma^2;k,p);s]}
  {1-\frac{8\lambda}{\sqrt{3}\pi}\;\left(-1\right)^l\;
  \calM[Q_l\left(x+\frac{1}{x}\right);s-1]}+
  \non\\&&
  \int\limits_{\sigma_+-\ii\infty}^{\sigma_++\ii\infty}
  \deint{}{s}p^{-s}\;
  \frac{8\pi\lambda\;\calM_+[\lim\limits_{k\to0}\calK^{(l)}
    (\frac{3k^2}{4}-\gamma^2;k,p);s]}
  {1-\frac{8\lambda}{\sqrt{3}\pi}\;\left(-1\right)^l\;
  \calM[Q_l\left(x+\frac{1}{x}\right);s-1]}+
  \\&&
  \oint\deint{}{s}p^{-s}\;
  \frac{S(p)}{1-\frac{8\lambda}{\sqrt{3}\pi}\;\left(-1\right)^l\;
  \calM[Q_l\left(x+\frac{1}{x}\right);s-1]}\;\bigg]\;\;,\non
\end{eqnarray}
where $\sigma_->\Re[l+2]$ and $\sigma_+<\Re[l+2]$. The denominator is simply
the Mellin transform of the resolvent of the Faddeev equation. Not
surprisingly, it determines the asymptotics of the solution. The function
$S(p)$, determined by the boundary-conditions, is in general an analytic
function in the strip $\sigma_-<\Re[p]<\sigma_+$ in which the integration
contour lies.

We thus see that the particular solution is finally defined everywhere except
at those points $\mathrm{Re}[s]=\mathrm{Re}[l]\pm2$ where
$\calM_{\pm}[\lim\limits_{k\to0}\calK^{(l)}]$ does not exist. It is not
necessary to perform the contour-integrations leading to an analytic solution
here. Rather, we note that
\begin{equation}
  \label{eq:ab}
  t_\lambda^{(l)}(p)=
  \sum\limits_{i=1}^\infty c_i\;\frac{k^l}{p^{s_l^{(i)}(\lambda)+1}}\;\;
  \text{ with }
  1=\frac{8\lambda}{\sqrt{3}\pi}\;\left(-1\right)^l\;
  \calM[Q_l\left(x+\frac{1}{x}\right);s_l^{(i)}(\lambda)]
\end{equation}
with some fixed coefficients $c_i$ with which the $i$th zero
$s_l^{(i)}(\lambda)$ of the denominator in \eqref{eq:aa} enters at fixed
$(l;\lambda)$ -- unfortunately, there is no closed form for these residues.
We used that because $Q_l(x+1/x)$ is real and symmetric under $x\to 1/x$, the
zeroes in the denominator of \eqref{eq:aa} come in quadruplets $\{\pm
s_l^{(i)}(\lambda);\pm s_l^{(i)\ast}(\lambda)\}$. Only the $s^{(i)}:=s$
closest to $-1$ is important for the amplitude at large $p$, as it provides
the strongest UV-dependence. Notice again that only those solutions exist
which do not diverge as $p\to\infty$ and for which the Mellin transformation
$\calM[Q_l\left(x+\frac{1}{x}\right);s]$ exists as well.

\subsection{How To Do an Integral}
\label{app:integral}

To obtain the zeroes of the denominator -- or equivalently equation
\eqref{eq:s} in the main text -- we now perform the Mellin transformation of
$Q_l\left(x+\frac{1}{x}\right)$. First, one represents the Legendre polynomial
by a hyperbolic function~\cite[eq.~(8.820.2)]{Gradstein}, and then uses in
turn the series-representation for ${}_2F_1$~\cite[eq.~(9.100)]{Gradstein}:
\begin{eqnarray}
  \label{eq:toseries}
  Q_l\left(x+\frac{1}{x}\right)&=&
  \frac{\sqrt{\pi}\;\Gamma[l+1]}{2^{l+1}\;\Gamma[l+\frac{3}{2}]}\;
  \left(x+\frac{1}{x}\right)^{-l-1}\;{}_2F_1[\frac{l+2}{2},\frac{l+1}{2};
  l+\frac{3}{2};\left(x+\frac{1}{x}\right)^{-2}]\\
  &=&\frac{\sqrt{\pi}\;\Gamma[l+1]}
  {2^{l+1}\;\Gamma[\frac{l}{2}+1]\;\Gamma[\frac{l+1}{2}]}\;
  \sum\limits_{n=0}^\infty \frac{\Gamma[\frac{l}{2}+1+n]\;
    \Gamma[\frac{l+1}{2}+n]}{\Gamma[l+\frac{3}{2}+n]\;\Gamma[n+1]}\;
  \left(x+\frac{1}{x}\right)^{-(2n+l+1)}\non
\end{eqnarray}
This series is convergent because $\left(x+\frac{1}{x}\right)^{-2}<1$ for all
$x\in[0;\infty]$, cf.~\cite[eq.~(9.102)]{Gradstein}. Now, perform the Mellin
transformation of each term using~\cite[eq.~(3.251.2)]{Gradstein}:
\begin{equation}
  \label{eq:toint}
  \int\limits_0^\infty\deint{}{x} x^{2n+s+l}\;(x^2+1)^{-(2n+l+1)}=
  \half\;\frac{\Gamma[n+\frac{l+s+1}{2}]\;\Gamma[n+\frac{l-s+1}{2}]}
  {\Gamma[2n+l+1]}
\end{equation}
This integral exists for $\Re[2n+l+1]>|\Re[s]|$, and hence for sufficiently
large $n$, i.e.~for an infinite, absolutely converging sequence.  By analytic
continuation, the result can thus be shown to be correct for all $n$. After a
few simple manipulations also with the aid of the doubling
formula~\cite[eq.~(8.335.1)]{Gradstein}, one can re-sum the series again:
\begin{eqnarray}
  \label{eq:tomellin}
  \calM[Q_l\left(x+\frac{1}{x}\right);s]&=&
  \sqrt{\pi}\;2^{-(l+2)}\sum\limits_{n=0}^\infty 4^{-n}\;
  \frac{\Gamma[n+\frac{l+s+1}{2}]\;\Gamma[n+\frac{l-s+1}{2}]}
  {\Gamma[n+l+\frac{3}{2}]\;\Gamma[n+1]}\\
  &=&
  \sqrt{\pi}\;2^{-(l+2)}\;
  \frac{\Gamma\left[\frac{l+s+1}{2}\right]\Gamma\left[\frac{l-s+1}{2}\right]}
  {\Gamma\left[\frac{2l+3}{2}\right]}\;
  {}_2F_1\left[\frac{l+s+1}{2},\frac{l-s+1}{2};
    \frac{2l+3}{2};\frac{1}{4}\right]\;\;.\non
\end{eqnarray}
Inserting this into \eqref{eq:ab} leads to the algebraic equation \eqref{eq:s}
for the coefficients $s_l(\lambda)$ given in the main text.

\absatz When does a solution to the homogeneous version of \eqref{eq:master}
exist? In general, no arbitrary homogeneous terms can be added due to
Fredholm's alternative: A non-zero solution exists for a given boundary
condition either for the inhomogeneous or for the homogeneous integral
equation. This follows also from the considerations leading to \eqref{eq:aa}
because the two regions in which
$\calM_-[\lim\limits_{k\to0}\calK^{(l)}(\frac{3k^2}{4}-\gamma^2;k,p);s]$ and
$\calM_+[\lim\limits_{k\to0}\calK^{(l)}(\frac{3k^2}{4}-\gamma^2;k,p);s]$ are
defined do in general not overlap, so that $S(p)$ has no support.

However, when $\calM[Q_l\left(x+\frac{1}{x}\right);s]$ itself exists, then the
homogeneous version of the integral equation has a solution. In that case,
\eqref{eq:toint} exists for each $n$, and in particular for $n=0$, so that one
must have $\Re[l+1]>|\Re[s]|$. As shown in Sect~\ref{sec:amplitude}, the
kernel is then singular, circumventing Fredholm's alternative.
Danilov~\cite{danilov} discussed the case $(l=0;\lambda=1)$, where the Mellin
transformation is listed in \cite[eq.~(4.296.3)]{Gradstein} with the
constraint $1>|\Re[s_{l=0}]|$, consistent with our result.

\absatz To summarise, all Mellin transformations of the particular solution in
\eqref{eq:aa} are well-defined for all $(s,l,\lambda)$ except for the driving
term, and for the back-transformation \eqref{eq:aa}. This constrains the
values of $s$ to:
\begin{equation}
  \label{eq:permits}
  \Re[s]\not=\mathrm{Re}[l]\pm2\;\;,\;\;\Re[s]>-1
\end{equation}
The homogeneous part of the Faddeev equation has in general a solution only if
the kernel is not compact. This is found for
\begin{equation}
  \label{eq:permithom}
 |\Re[s]|<\Re[l+1]\;\;.
\end{equation}

\subsection{The Full Off-shell Amplitude}
\label{app:offshell}

One obtains the solution to the full off-shell Faddeev equation
\eqref{eq:offshell} easily as follows: Replace $ 8\pi\lambda\;
\calM_\pm[\lim\limits_{k\to0}\calK^{(l)}(\frac{3k^2}{4}-\gamma^2;k,p);s]$ in
App.~\ref{app:construction} by
\begin{eqnarray}
 \calM[8\pi\lambda\;\frac{(-1)^l}{kp}\;
 Q_l\left(\frac{p}{k}+\frac{k}{p}\right);s]=
 8\pi\lambda\;\left(-1\right)^l\;k^{s-2}\;
 \calM[Q_l\left(x+\frac{1}{x}\right); s-1]\;\;.
\end{eqnarray}  
The asymptotic form given in \eqref{eq:offshellsol} follows now from the
analogue to (\ref{eq:aa}/\ref{eq:ab}), keeping in mind that the contours can
only be closed in the positive half-plane when $k<p$, and in the negative one
when $k>p$. Notice that \emph{both} off-shell momenta must obey the integral
equations \eqref{eq:offshell}. 

\newpage






\begin{thebibliography}{99}
  
\bibitem{NDA} A.~Manohar and H.~Georgi, \NPB\textbf{234} (1984), 189
  <n.b.~Acknowledgement>; H.~Georgi and L.~Randall, \NPB\textbf{276} (1986),
  241.
  
\bibitem{Weinberg} S.~Weinberg, Nucl.~Phys.~B \textbf{363}, 3 (1991).
    
  
\bibitem{bira_review} U.\ van Kolck, Prog.\ Part.\ Nucl.\ Phys.\ \textbf{43},
  337 (1999) [nucl-th/9902015].
  
\bibitem{seattle_review} S.~R.\ Beane, P.~F.\ Bedaque, W.~C.\ Haxton, D.~R.\ 
  Phillips and M.~J.\ Savage, in ``At the frontier of particle physics'', M.\ 
  Shifman (ed.), World Scientific, 2001 [nucl-th/0008064].
  
\bibitem{bedaque_bira_review}
  P.~F.~Bedaque and U.~van Kolck,
  Ann.\ Rev.\ Nucl.\ Part.\ Sci.\ {\bf 52}, 339 (2002) [nucl-th/0203055].

\bibitem{3stooges_doublet} P.~F.\ Bedaque, H.-W.\ Hammer and U.\ van Kolck,
  Nucl.\ Phys.\ A \textbf{676}, 357 (2000) [nucl-th/9906032].

\bibitem{doubletNLO} H.-W.\ Hammer and T.\ Mehen, Phys.\ Lett.\ B
  \textbf{516}, 353 (2001) [nucl-th/0105072].
  
\bibitem{4stooges} P.~F.~Bedaque, G.~Rupak, H.~W.~Grie\3hammer and
  H.-W.~Hammer,
  Nucl.\ Phys.\ A {\bf 714}, 589 (2003) [nucl-th/0207034].

\bibitem{Sadeghi:2004es} H.~Sadeghi and S.~Bayegan,
  Nucl.\ Phys.\ A {\bf 753}, 291 (2005) [nucl-th/0411114].
  
\bibitem{danilov} G.~S.\ Danilov, Sov.\ Phys.\ JETP \textbf{13}, 349 (1961).
  
\bibitem{3stooges_boson} P.~F.\ Bedaque, H.-W.\ Hammer and U.\ van Kolck,
  Phys.\ Rev.\ Lett.\ \textbf{82}, 463 (1999) [nucl-th/9809025]; Nucl.\ Phys.\ 
  A {\bf 646}, 444 (1999) [nucl-th/9811046].
  
\bibitem{wilson} K.~G.~Wilson, \PRD\textbf{3}, 1818 (1971); S.~D.~G{\l}azek
  and K.~G.~Wilson, \PRD\textbf{47}, 4657 (1993); S.~D.~G{\l}azek and
  K.~G.~Wilson, Phys.\ Rev.\ Lett.\ {\bf 89}, 230401 (2002) [hep-th/0203088].

\bibitem{Braaten:2003eu} E.~Braaten and H.~W.~Hammer,
  Phys.\ Rev.\ Lett.\ \textbf{91}, 102002 (2003) [nucl-th/0303038].

\bibitem{efimovI} V.\ Efimov, Nucl.\ Phys.\ A \textbf{362}, 45 (1981); Phys.\ 
  Rev. C \textbf{44}, 2303 (1991); V.\ Efimov and E.~G.\ Tkachenko, Phys.\ 
  Lett.\ B \textbf{157}, 108 (1985).

\bibitem{Afnan:2003bs} I.~R.~Afnan and D.~R.~Phillips,
  Phys.\ Rev.\ C {\bf 69}, 034010 (2004) [nucl-th/0312021].
  
 \bibitem{BarfordBirse} T.~Barford,
   PhD-thesis, Manchester University 2004, [nucl-th/0404072]; T.~Barford and
   M.~C.~Birse, J.\ Phys.\ A~\textbf{38}, 697 (2005)
   [nucl-th/0406008].

\bibitem{Braaten:2004rn} E.~Braaten and H.~W.~Hammer,
  cond-mat/0410417.
  
\bibitem{suppressed3bfs} H.~W.~Grie\3hammer, in: Mini-proceedings of ``Chiral
  Dynamics: Theory and Experiment (CD2003),'' eds.~U.-G.~Mei\3ner,
  H.-W.~Hammer and A.~Wirzba, hep-ph/0311212.
  
\bibitem{improve3body} H.~W.~Grie{\ss}hammer, \NPA\textbf{744} (2004), 192
  [nucl-th/0404073].

\bibitem{Kaplan:1996nv} In EFT, this was first considered by D.~B.~Kaplan,
  Nucl.\ Phys.\ B {\bf 494}, 471 (1997) [nucl-th/9610052].
  
\bibitem{2stooges_quartet} P.~F.\ Bedaque and U.\ van Kolck, Phys.\ Lett.\ B
  \textbf{428}, 221 (1998) [nucl-th/9710073].
  
\bibitem{3stooges_quartet} P.~F.\ Bedaque, H.-W.\ Hammer and U.\ van Kolck,
  Phys.\ Rev.\ C \textbf{58}, R641 (1998) [nucl-th/9802057].
  
\bibitem{pbhg} P.~F.\ Bedaque and H.~W.\ Grie{\ss}hammer, Nucl.\ Phys.\ A {\bf
    671}, 357 (2000) [nucl-th/9907077].
  
\bibitem{Bethe} J.~Schwinger, hectographed notes on nuclear physics, Harvard
  University 1947; G.~F.~Chew and M.~L.~Goldberger, Phys.\ Rev.\ \textbf{75},
  1637 (1949); F.~C.~Barker and R.~E.~Peierls, Phys.\ Rev.\ \textbf{75}, 3122
  (1949); H.~A.~Bethe, Phys.\ Rev.\ \textbf{76}, 38 (1949).
  
\bibitem{skorny} G.~V.~Skorniakov and K.~A.~Ter-Martirosian, Sov.\ Phys.\ JETP
  \textbf{4}, 648 (1957).
  
\bibitem{chickenpaper} F.~Gabbiani, P.~F.\ Bedaque and H.~W.\ Grie{\ss}hammer,
  Nucl.\ Phys.\ A \textbf{675}, 601 (2000) [nucl-th/9911034].
  
\bibitem{Gradstein} I.\ S.\ Gradshteyn and I.\ M.\ Ryzhik, Table of Integrals,
  Series and Products, 5th edition, Academic Press, San Diego 1994.
  
\bibitem{su4} E.\ Wigner, Phys.\ Rev.\ \textbf{51}, 106 (1939); Phys.\ Rev.\ 
  \textbf{51}, 947 (1939); Phys.\ Rev.\ \textbf{56}, 519 (1939).
  
\bibitem{Mehen:1999qs}
T.~Mehen, I.~W.~Stewart and M.~B.~Wise,
Phys.\ Rev.\ Lett.\ \textbf{83}, 931 (1999) [hep-ph/9902370].

\bibitem{thomas} L.~W.~Thomas, Phys.~Rev.~\textbf{47}, 903 (1935).
  
\bibitem{phillips} A.~C.\ Phillips and G.\ Barton, Phys.\ Lett.\ B
  \textbf{28}, 378 (1969).
  
\bibitem{tkachenko} V.\ Efimov and E.~G.\ Tkachenko, Few-Body Syst.\ 
  \textbf{4}, 71 (1988).
  
\bibitem{Lepage:1997cs} G.~P.~Lepage, ``How to renormalize the Schr\"odinger
  equation'', lectures given at 9th Jorge Andre Swieca Summer School:
  Particles and Fields, Sao Paulo, Brazil, 16-28 Feb 1997, nucl-th/9706029.

\bibitem{Bernard:2003rp} V.~Bernard, T.~R.~Hemmert and U.~G.~Mei\3ner,
  Nucl.\ Phys.\ A \textbf{732},149 (2004) [hep-ph/0307115].

\bibitem{Epelbaum:2003gr} E.~Epelbaum, W.~Gloeckle and U.~G.~Mei\3ner,
  Eur.\ Phys.\ J.\ A \textbf{19}, 125 (2004) [nucl-th/0304037].
  
\bibitem{boundcoherentexp} T.~C.~Black et al., \PRL~\textbf{90}, 192502
   (2003); K.~Sch\"on et al., \PRC~\textbf{67}, 044005 (2003).
  
\bibitem{doublet_sca} W.\ Dilg, L.\ Koester and W.\ Nistler, Phys.\ Lett.\ B
  \textbf{36}, 208 (1971).

\bibitem{Friar:1999jd} J.~L.~Friar, D.~H\"uber, H.~Wita{\l}a and G.~L.~Payne,
  Acta Phys.\ Polon.\ B {\bf 31}, 749 (2000) [nucl-th/9908058].
  
\bibitem{Witala:2003mr} H.~Wita{\l}a, A.~Nogga, H.~Kamada, W.~Gl\"ockle,
  J.~Golak and R.~Skibinski,
  nucl-th/0305028.
  
  
\bibitem{Blankleider:2000vi} B.~Blankleider and J.~Gegelia,
  [nucl-th/0009007].

\bibitem{Griesshammer:2000mi} H.~W.~Grie\3hammer and G.~Rupak,
  \journal{\PLB}{529}{2002}{57} [nucl-th/0012096].

\bibitem{ItzyksonZuber} C.~Itzykson and J.-B.~Zuber, Quantum Field
  Theory, McGraw-Hill, New York et al.~1980.
  
\bibitem{Huber:1998hu} see e.g.~D.~H\"uber and J.~L.~Friar,
  Phys.\ Rev.\ C {\bf 58}, 674 (1998) [nucl-th/9803038].
  
\bibitem{Beane:2001bc} S.~R.~Beane, P.~F.~Bedaque, M.~J.~Savage and U.~van
  Kolck,
  Nucl.\ Phys.\ A \textbf{700}, 377 (2002) [nucl-th/0104030].
  
\bibitem{nogga} A.~Nogga, in: Mini-proceedings of the \textsc{337th WE Heraeus
    Seminar: Effective Field Theories in Nuclear, Particle, and Atomic
    Physics}, Bad Honnef (Germany), 15th December 2004, eds.~J.~Bijnens,
  U.-G.~Mei\3ner and A.~Wirzba, hep-ph/0502008; A.~Nogga, R.~G.~E.~Timmermans
  and U.~van Kolck, nucl-th/0506005.

\bibitem{MorseFeshbach} P.~M.~Morse and H.~Feshbach, Methods of Theoretical
  Physics, Part I, McGraw-Hill, New York et al.~1953.

\end{thebibliography}
\end{document}